\documentclass[journal]{IEEEtran}
\usepackage{amsmath,amssymb,amsfonts,amsthm}
\usepackage{graphicx}
\usepackage[caption=false,font=footnotesize]{subfig}
\usepackage{cite}
\usepackage{booktabs}
\usepackage{placeins}
\usepackage[hidelinks]{hyperref}
\graphicspath{{figs/}}
\newtheorem{proposition}{Proposition}
\newtheorem{remark}{Remark}
\newcommand{\bv}[1]{\mathbf{#1}}
\newcommand{\E}{\mathbb{E}}
\DeclareMathOperator{\diag}{diag}
\DeclareMathOperator{\tr}{tr}

\begin{document}
\title{\huge Movable-Element STAR-RIS and Rate-Splitting Multiple Access for Wideband Terahertz Industrial IoT Networks}
\author{Wali~Ullah~Khan,~\IEEEmembership{Member,~IEEE}}
\maketitle

\begin{abstract}
A simultaneously transmitting and reflecting reconfigurable intelligent surface (STAR-RIS) can extend terahertz (THz) coverage to industrial devices on both sides of a partition. A dominant base-station (BS) path to the surface can, however, make the effective device channels nearly collinear at the BS and limit spatial multiplexing. We study an 850-GHz industrial Internet of Things downlink with a hybrid BS, rate-splitting multiple access (RSMA), and a STAR-RIS composed of movable tiles. We derive conditional channel moments under actuator error, vibration, and device-location uncertainty and characterize the BS-side subspace created by the common propagation paths. The design precomputes surface states and precoders for candidate device groups, assigns whole coherence blocks to the groups, and optimizes tile placement on a slower timescale. In the six-device factory study, space-division multiple access with time sharing increases mean delivered sum throughput by 36\% and 40\% over one-slot space-division multiple access at 30 and 40~dBm, respectively, under the evaluated outage calibration and switching assumptions. RSMA and space-division multiple access have similar ergodic sum rates under the evaluated schedules. Tile movement changes mean delivered sum throughput by less than 1\% for RSMA with time sharing and per-slot STAR states. The minimum-rate outage results show how dense clutter limits the service available to industrial devices.
\end{abstract}


\section{Introduction}\label{sec:intro}
Wideband terahertz (THz) communication offers substantial bandwidth for industrial Internet of Things (IIoT) downlinks, including high-resolution operator visualization and the delivery of digital-twin updates~\cite{rappaport2019,akyildiz2014}. An industrial network may also serve motion-control devices whose operation depends on timely and reliable packet delivery~\cite{3gpp22104}. Supporting these traffic classes in the same factory requires both high-capacity links and reliable coverage across production cells. At THz frequencies, the link budget relies on directional transmission, while metal machinery and safety partitions can obstruct the resulting propagation paths. Maintaining coverage is therefore a central challenge, even when ample bandwidth is available.

A simultaneously transmitting and reflecting reconfigurable intelligent surface (STAR-RIS) mounted in a safety partition can redirect energy toward devices in both adjacent production cells~\cite{mu2022star}. A movable-element STAR-RIS (ME-STAR-RIS) adds a geometric degree of freedom through tile repositioning. Changing the tile positions alters the relative propagation delays across the aperture and can improve wideband focusing for a given device geometry~\cite{zhu2024ma,zhu2025mestarsnf}. At submillimeter wavelengths, actuator errors and machinery-induced vibration can also perturb those delays and reduce coherent combining. The value of tile movement therefore depends on the propagation geometry, the available electronic surface states, and the accuracy with which the surface can be positioned.

Extending coverage to both sides of the partition can still leave the surface-served devices with nearly collinear effective channels at the base station (BS) when they share a dominant BS--surface path. In that regime, spatially separating private streams is difficult. Rate-splitting multiple access (RSMA) offers a way to manage the resulting interference by allowing the BS to encode part of the traffic in a common stream that the scheduled devices decode before recovering their private streams~\cite{clerckx2023primer,joudeh2016robust}. Scheduling device groups in different coherence blocks can also reduce interference and allow the STAR-RIS to focus on the active group. These options motivate the question studied here: how do rate splitting, group scheduling, per-slot surface configuration, and tile repositioning affect delivered throughput and minimum-rate service in a wideband THz factory downlink?

\subsection{Related Work}
Industrial channel measurements establish that factory layout and metal infrastructure affect both coverage and the structure of the received multipath. Wang et al.~\cite{wang2025factory} measured four indoor-factory settings at 28, 38, and 132~GHz and developed a geometry-based stochastic model from the measured channels. Ju and Rappaport~\cite{ju2024statistical} derived wideband channel statistics from line-of-sight (LoS) and non-LoS links measured at 142~GHz in four factories, extending earlier single-factory measurements~\cite{ju2022factory}. More recently, Shao et al.~\cite{shao2026factory} characterized path loss, delay spread, and angular spread at 318~GHz in a factory hall and a warehouse. Together, these studies motivate modeling both link blockage and scattered propagation. The 850-GHz model uses frequency-specific gaseous attenuation and extrapolates factory blockage beyond those measurement bands.

Wideband THz links also depend on how the aperture responds across frequency. Dai et al.~\cite{dai2022dpp} used delay-phase precoding to reduce beam split in large arrays with frequency-flat phase shifters. At the surface, Dutin et al.~\cite{dutin2025surface} measured the frequency-dependent insertion loss of a passive reflector around 300~GHz and demonstrated a non-LoS data link. This experiment illustrates the dependence of a redirected link on the frequency response of the surface. For passive STAR operation, Xu et al.~\cite{xu2022coupled} established the coupling between transmission and reflection phases. The present model combines this passivity constraint with a shared, state-dependent frequency response across the two sides of the surface.

\begin{figure*}[!t]
\centering
\includegraphics[width=\textwidth]{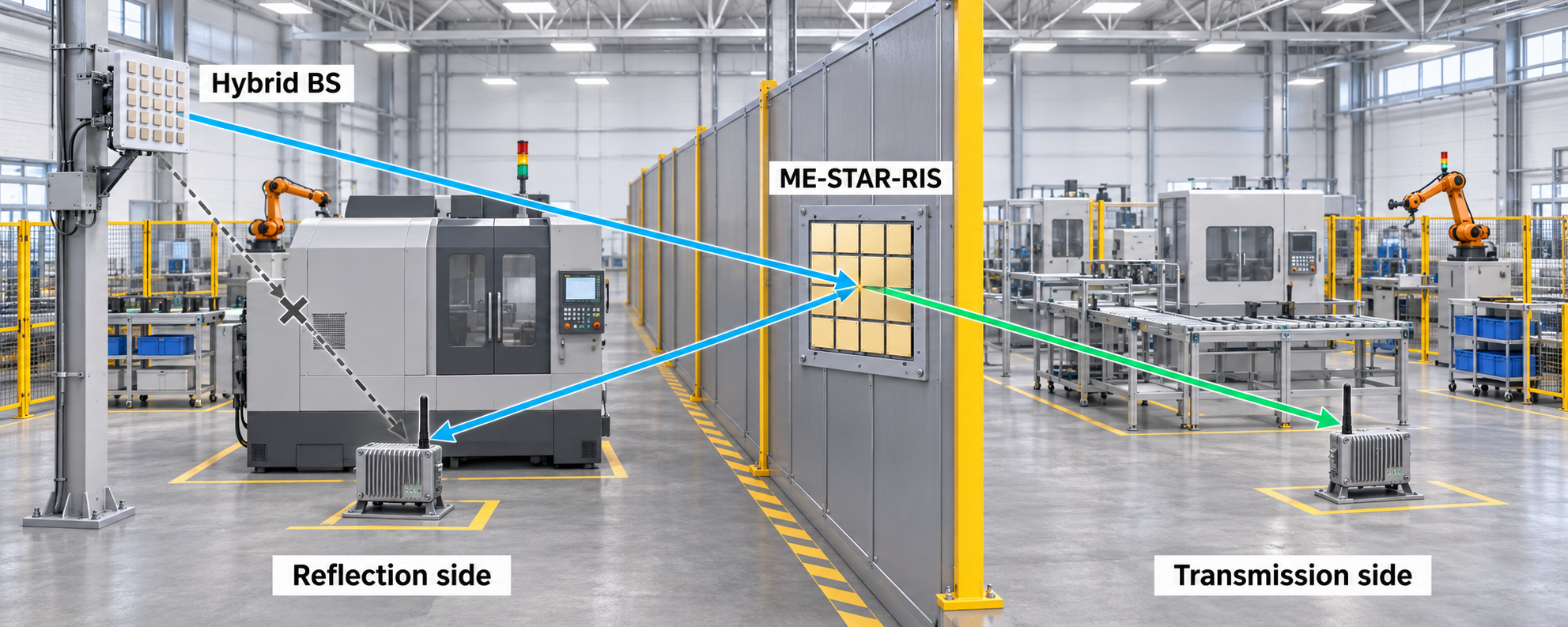}
\caption{Downlink across a factory partition. The BS illuminates a tile-based ME-STAR-RIS, which serves single-antenna devices on the reflection and transmission sides. Device locations and geometry are schematic.}
\label{fig:deployment}
\end{figure*}

A movable aperture introduces position control alongside electronic beamforming~\cite{zhu2024ma}. Zheng et al.~\cite{zheng2025twotime} studied two-timescale resource allocation for multiuser movable-antenna MIMO. Zhao et al.~\cite{zhao2024mestars} jointly optimized element positions, transmit precoders, and surface coefficients under several STAR protocols, reporting stronger movement gains in simulated settings with more users or scatterers. Guangyu Zhu et al.~\cite{zhu2025mestarsnf} compared four motion constraints for mitigating near-field wideband beam squint with movable STAR elements. Yanze Zhu et al.~\cite{zhu2025movsub} used movable BS antennas and reflecting subarrays to mitigate beam squint on a wideband THz link. These studies motivate testing whether the gains from repositioning persist when the surface already has discrete electronic phase control and when its tiles experience common and differential motion.

The spatial rank of the BS--surface hop also affects which devices can be served together. Nadeem et al.~\cite{nadeem2020rank} analyzed max-min SINR for reflecting-RIS systems with rank-one and full-rank BS--RIS channels, establishing the penalty associated with a rank-one hop. Salem et al.~\cite{salem2025clustering} showed how user clustering and surface configuration can be optimized together in a full-duplex STAR-RIS system using nonorthogonal multiple access. These results motivate examining both the BS-side spatial subspace and the allocation of transmission time among device groups in a partition-mounted STAR-RIS downlink.

RSMA manages interference by splitting each message into components that can be decoded jointly through a common stream or separately through private streams~\cite{mao2018rsma,clerckx2023primer,joudeh2016robust}. Zhang et al.~\cite{zhang2025rsma} studied near-field RSMA with imperfect transmitter channel knowledge and residual interference after successive interference cancellation. Zhou et al.~\cite{zhou2026robust} developed a robust multiuser MIMO precoder using a generalized mutual-information bound, and Asif et al.~\cite{asif2026mestarsrsma} jointly optimized RSMA precoding, rate allocation, STAR coefficients, and element positions under channel uncertainty. These works establish closely related precoding and surface-design methods. The question here is how the common stream, time sharing, and slot-specific surface states compare when the same industrial downlink is evaluated using scheduled rates, receiver pilots, and minimum-rate service outage.

\subsection{Motivation and Contributions}
An unresolved deployment question is how to allocate resources when a single partition-mounted STAR-RIS provides coverage to both production cells but illuminates nearly the same BS spatial mode for every surface-served device. Mechanical repositioning can change propagation delays and device-side focusing, while RSMA can change how interference is decoded. Their benefits depend on the propagation geometry, the availability of time sharing and the rates that receivers can actually decode after channel estimation. We study these mechanisms within one downlink and make the following contributions.
\begin{itemize}
\item We develop an 850-GHz wideband IIoT downlink model with a hybrid BS and a partition-mounted STAR-RIS composed of movable tiles. The model captures spherical propagation across tiles, atmospheric attenuation, factory scattering and blockage, a passive dispersive STAR response, and pilot tracking of the effective channels.
\item We derive conditional first and second channel moments under actuator error, common and differential vibration, and device-location uncertainty. We show how common phase rotation and differential phase error affect the channel moments differently. We also characterize the BS-side spatial subspace generated by the shared paths and identify when these paths yield a dominant common spatial direction.
\item We formulate scheduled-rate outage and minimum-rate shortfall and develop a two-timescale design. The electronic stage constructs candidate device groups with discrete STAR states and hybrid precoders and allocates time by linear programming. The mechanical stage searches feasible tile positions. The evaluation uses a six-device factory scenario and held-out layouts for scheduled-rate delivery. It compares scheduling, per-slot STAR states, multiple access, and tile movement under the same deployment assumptions.
\end{itemize}

Section~\ref{sec:model} presents the channel, hardware and receiver models. Sections~\ref{sec:problem} and~\ref{sec:solution} formulate the design problem and describe the solution. Section~\ref{sec:results} presents the full-system evaluation, and Section~\ref{sec:conclusion} concludes the paper.

Throughout the paper, bold lowercase and uppercase letters denote vectors and matrices. Transpose, complex conjugation, and Hermitian transpose are denoted by $(\cdot)^T$, $(\cdot)^*$, and $(\cdot)^H$, respectively. The Euclidean and Frobenius norms are $\|\cdot\|$ and $\|\cdot\|_F$. We use $\E[\cdot]$ for expectation, $\tr(\cdot)$ for the trace, and $\mathbb{1}\{\cdot\}$ for the indicator of an event. The distributions $\mathcal{N}$ and $\mathcal{CN}$ are real Gaussian and proper complex Gaussian, respectively. Channel coefficients are transfer gains, so the cascaded channels use ordinary transposes and the propagation phases of successive hops add.

\section{System Model}\label{sec:model}
\subsection{Factory Deployment and Timescales}
An ME-STAR-RIS is mounted in a partition between two production cells, as shown in Fig.~\ref{fig:deployment}. The surface lies in the $x$--$y$ plane of a global coordinate system, with its normal $\hat{\bv{z}}$ pointing into the reflection half-space $\mathcal{H}_{\rm R}=\{z>0\}$, which contains the BS. The BS has a uniform planar array (UPA) of $N_t$ antennas at half-wavelength spacing and $N_{\rm RF}$ radio-frequency chains. It serves $K$ single-antenna devices, $\mathcal{K}=\mathcal{K}_{\rm R}\cup\mathcal{K}_{\rm T}$, where $\mathcal{K}_{\rm R}$ ($\mathcal{K}_{\rm T}$) contains the devices in $\mathcal{H}_{\rm R}$ ($\mathcal{H}_{\rm T}=\{z<0\}$). We write $\chi_k\in\{{\rm R},{\rm T}\}$ for the side of device~$k$. By traffic class, $\mathcal{K}=\mathcal{K}_{\rm B}\cup\mathcal{K}_{\rm G}$: best-effort high-rate downlink devices (operator XR headsets, digital-twin and AR model synchronization, AGV map updates) in $\mathcal{K}_{\rm B}$, and mission-critical devices (robot and motion controllers) in $\mathcal{K}_{\rm G}$, each with a packet size $D_k$, a deadline $T_k$ and an outage target $\epsilon_k$ that describe its service profile~\cite{3gpp22104}. The simulations use a parameterized rate threshold for these devices. The model is downlink only; uplink traffic such as machine-vision cameras is part of the wider motivation. Devices in $\mathcal{H}_{\rm T}$ have no direct link to the BS because the partition blocks it.

Transmission uses orthogonal frequency-division multiplexing (OFDM) with bandwidth $B$ around the carrier $f_c$. Here $c$ denotes the speed of light. We evaluate the band at $M$ frequency samples $f_m=f_c+\left(m-\frac{M+1}{2}\right)\frac{B}{M}$, where $m=1,\ldots,M$. The wavenumber is $k_m=2\pi f_m/c$ and the carrier wavelength is $\lambda_c=c/f_c$. These samples are quadrature points for band-averaged rates, not physical tones; the physical OFDM numerology determines cyclic-prefix and pilot overhead in Section~\ref{sec:rates}.

The design separates mechanical placement from the computation and execution of electronic configurations. Tile positions are optimized on the timescale of changes in the factory layout and are fixed during a scheduling frame. For each estimated layout and tile placement, the BS computes candidate STAR states, precoders, and rate splits before transmission and reuses them across the frame. Each coherence block serves one selected device group using the corresponding precomputed configuration. Receivers use pilots in each block to track the residual effective-channel variation. At 850~GHz, a device moving at 2~m/s has a maximum Doppler shift of approximately 5.7~kHz. We use 70~$\mu$s as the reference block duration in the pilot accounting.

\subsection{Tile-Based ME-STAR-RIS}
At 850~GHz, moving each of several thousand elements independently with micrometre precision is not realistic. We therefore group the elements into $Q$ rigid tiles, each carrying $S\times S$ elements at a fixed spacing $d_e=\lambda_c/2$, so the tile width is $W=Sd_e$ and the surface has $N=QS^2$ elements. Each tile has an independent actuator, such as a piezoelectric or voice-coil stage. The model assumes millimeter-scale stroke and micrometer-scale positioning errors. Tile $q$ is translated to the centre position $\bv{p}_q=[x_q,y_q,z_q]^T$. Element $n$ of tile $q$ sits at
\begin{equation}
\bv{e}_{q,n}=\bv{p}_q+\bv{r}_n,\qquad \bv{r}_n=[\,\xi_n d_e,\ \upsilon_n d_e,\ 0\,]^T,
\end{equation}
with $\xi_n,\upsilon_n\in\{-\tfrac{S-1}{2},\dots,\tfrac{S-1}{2}\}$. The tiles translate but do not rotate. The feasible set of the positions $\bv{P}=[\bv{p}_1,\dots,\bv{p}_Q]$ is
\begin{align}
\mathcal{P}=\Big\{\bv{P}:\ & |x_q|,|y_q|\le \tfrac{A-W}{2},\ |z_q|\le Z,\ \forall q,\nonumber\\
& \max(|x_q-x_{q'}|,|y_q-y_{q'}|)\ge W+g,\ \forall q\neq q'\Big\},\label{eq:Pset}
\end{align}
where $A\times A$ is the in-plane movement region, $Z$ is the maximum magnitude of out-of-plane displacement, and $g\ge 0$ is a guard gap that prevents the tiles from touching. With $S=1$, each tile comprises a single element and $W+g$ is the minimum center-to-center separation required along at least one in-plane axis.

The three movement directions have different physical roles. In-plane spreading enlarges the effective aperture, which increases the near-field region and the spatial resolution for separating devices, but it also increases the spread of path lengths across the aperture and hence beam split. Normal displacement $z_q$ changes the cascaded path length through tile~$q$ by approximately $-(\hat{u}_z+\hat{w}_z)z_q$, where $\hat{\bv{u}}$ and $\hat{\bv{w}}$ are the unit vectors from the tile towards the BS and towards the device. It therefore acts as a frequency-independent true-time delay, unlike the phase shift of an element. For a reflection-side device $\hat{u}_z+\hat{w}_z=\cos\vartheta_{\rm i}+\cos\vartheta_{\rm o}$ is large, while for a transmission-side device $\hat{u}_z+\hat{w}_z=\cos\vartheta_{\rm i}-\cos\vartheta_{\rm o}$ is small, so normal movement is a strong delay control for reflection-side devices and a weak one for transmission-side devices whose incidence and departure angles are similar (it is not zero for asymmetric transmission geometries).

\begin{figure*}[!t]
\centering
\subfloat[Specific attenuation; shaded: the chosen 835--865~GHz band]{\includegraphics[width=0.47\textwidth]{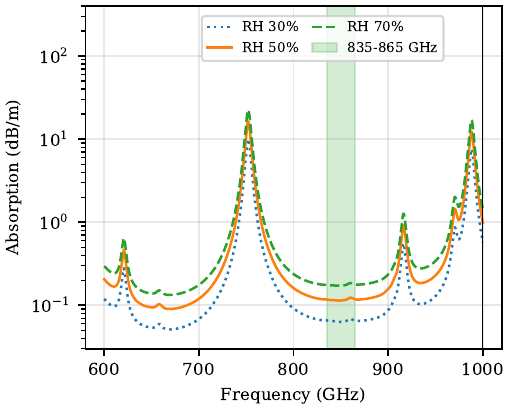}\label{fig:absorptiona}}\hfill
\subfloat[Absorption loss over 20~m across the band]{\includegraphics[width=0.47\textwidth]{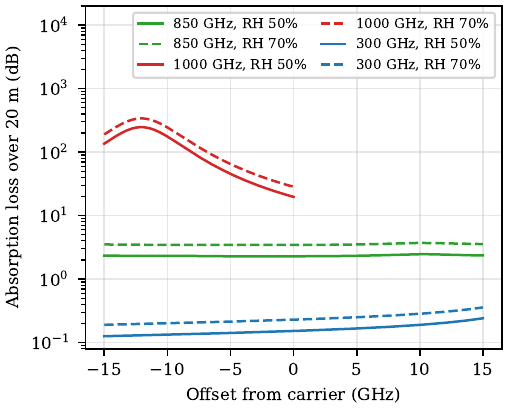}\label{fig:absorptionb}}
\caption{Molecular absorption from the ITU-R P.676-12 line-by-line model at 25~$^\circ$C and 1~atm. (a) Specific attenuation and the chosen band. (b) Absorption loss over a 20~m path across a 30~GHz band centred at 850~GHz and 300~GHz, and over 985--1000~GHz for the band centred at 1~THz (the model is valid only up to 1~THz).}
\label{fig:abs}
\end{figure*}

\subsection{Frequency-Dependent Passive STAR Element}
Most STAR-RIS studies assume transmission and reflection coefficients that are the same on every subcarrier. A practical tunable element is a resonator, and over a band of a few percent its phase drifts by an amount that depends on its configured state~\cite{abeywickrama2020}. We model each element as a symmetric two-port with a finite set $\mathcal{S}$ of states (the codebook). State $s=(\beta_s,\phi_s,\varsigma_s)$ has an energy split $\beta_s\in(0,1)$, a phase $\phi_s$ at $f_c$, and a sign $\varsigma_s\in\{+1,-1\}$. Its reflection and transmission responses share one dispersive factor:
\begin{equation}
r_s(f)=\sqrt{\eta\beta_s}\,\Gamma(f;\phi_s),\quad t_s(f)=j\varsigma_s\sqrt{\eta(1-\beta_s)}\,\Gamma(f;\phi_s),\label{eq:star}
\end{equation}
where $\eta\le1$ is the ohmic efficiency and $\Gamma$ is the all-pass response of a lossless resonator with loaded quality factor $Q_{\rm e}$, tuned so that its phase at $f_c$ equals $\phi$:
\begin{equation}
\Gamma(f;\phi)=e^{\,j\left(\pi-2\arctan x(f)\right)},\quad x(f)=2Q_{\rm e}\frac{f-f_{\rm r}}{f_{\rm r}},\label{eq:lorentz}
\end{equation}
with resonance $f_{\rm r}=f_c/\big(1+\cot(\phi/2)/(2Q_{\rm e})\big)$. Because $r_s^*t_s+t_s^*r_s=0$ and $|r_s|^2+|t_s|^2=\eta$ at every frequency, the scattering matrix $\bv{S}_s(f)=\big[\begin{smallmatrix}r_s&t_s\\t_s&r_s\end{smallmatrix}\big]$ satisfies $\bv{S}_s^H(f)\bv{S}_s(f)=\eta\bv{I}\preceq\bv{I}$ across the whole band. The element is therefore passive for any two-sided excitation, and the transmission phase stays in quadrature with the reflection phase on every subcarrier, as required for a lossless symmetric element~\cite{xu2022coupled}. The group delay is shared by both sides and depends on the state:
\begin{equation}
\tau_{\rm g}(\phi)=-\frac{1}{2\pi}\frac{\partial \angle\Gamma}{\partial f}\Big|_{f_c}=\frac{2Q_{\rm e}}{\pi f_{\rm r}}\sin^2\!\frac{\phi}{2},
\end{equation}
so states near resonance ($\phi\approx\pi$) are the most dispersive. The default codebook has 2-bit phases $\phi\in\{45^\circ,135^\circ,225^\circ,315^\circ\}$ (offset to avoid the singular state $\phi=0$), three split levels $\beta\in\{0.25,0.5,0.75\}$ and both signs, so $|\mathcal{S}|=24$. The values $Q_{\rm e}=10$ and $\eta=-1$~dB are assumed hardware parameters, not measured ones. The codebook is a lookup table, so a measured or full-wave simulated response can replace~\eqref{eq:lorentz}; a model with different reflection and transmission dispersion can be used too, provided its joint scattering matrix is checked for $\bv{S}_s^H\bv{S}_s\preceq\bv{I}$ at every frequency. We write $s_{q,n}\in\mathcal{S}$ for the state of element $(q,n)$, $\theta^{\rm R}_s=r_s$, $\theta^{\rm T}_s=t_s$, and $\bv{\Theta}_q^{\chi}[m]=\diag\big(\theta^{\chi}_{s_{q,1}}(f_m),\dots,\theta^{\chi}_{s_{q,S^2}}(f_m)\big)$.

\begin{figure*}[!t]
\centering
\subfloat[Maximum band-edge phase error versus configured phase]{\includegraphics[width=0.45\textwidth]{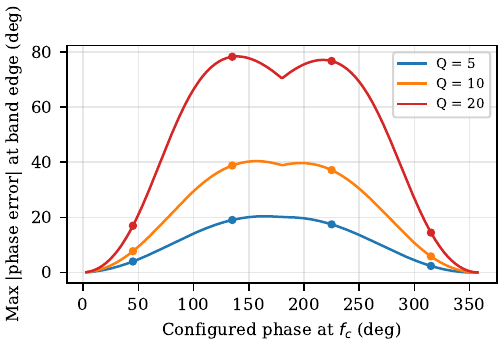}\label{fig:starresponsea}}\hfill
\subfloat[Phase drift of the codebook states, $Q_{\rm e}=10$]{\includegraphics[width=0.45\textwidth]{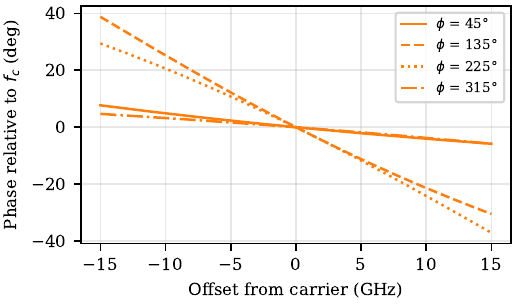}\label{fig:starresponseb}}
\caption{Shared dispersive factor $\Gamma$ of the passive STAR element~\eqref{eq:star}, which applies to both reflection and transmission: maximum band-edge phase error versus configured phase, with the 2-bit codebook phases marked (a), and phase drift relative to $f_c$ of the codebook states (b), $Q_{\rm e}=10$ unless stated, 850~GHz, 30~GHz bandwidth.}
\label{fig:star}
\end{figure*}

Figure~\ref{fig:star} shows the phase response of the element model. The phase drift varies among the codebook states, so even with a shared reflection and transmission response within each state, the relative phases between elements assigned different states can change across the band.

\subsection{Wideband Near-Field Channel Model}
\subsubsection{Molecular absorption and the choice of carrier}
A path of length $d$ at frequency $f$ is attenuated in power by $e^{-\kappa(f)d}$, where $\kappa(f)=\frac{\ln 10}{10}\gamma(f)$ and $\gamma(f)$ [dB/m] is the specific attenuation of the ITU-R P.676-12 line-by-line model~\cite{itu676}. This model sums 44 oxygen and 35 water-vapour resonance lines with pressure-broadened line shapes and a pseudo-line at 1780~GHz for the water-vapour continuum; it is valid up to 1~THz and depends on pressure, temperature and humidity. The simulation uses this revision of the ITU model for consistency with the absorption data. Molecular spectroscopic databases such as HITRAN provide a reference for alternative line-by-line calculations~\cite{hitran2020}. Fig.~\ref{fig:abs} shows $\gamma(f)$ at 25~$^\circ$C, 1~atm and relative humidity (RH) 30--70\%, typical of a factory hall.

Near 1~THz the 988~GHz water line dominates: at RH~50\% the absorption is 0.99~dB/m at exactly 1~THz and up to 12.4~dB/m within 985--1000~GHz, so the lower half of a 30~GHz band centred at 1~THz (the part inside the model's validity range) would lose between 20 and about 250~dB over a 20~m path. Between the 752 and 916~GHz water lines there is a transmission window. In 835--865~GHz the absorption is flat, between 0.114 and 0.124~dB/m at RH~50\%, or about 2.3~dB over 20~m (3.5~dB at RH~70\%). We therefore choose $f_c=850$~GHz and $B=30$~GHz. A second, slightly lower-loss window lies at 650--690~GHz (about 0.09~dB/m). Absorption is thus not the dominant loss at 850~GHz over factory distances (spreading loss is), but it varies with humidity, so we sweep RH in Section~\ref{sec:results}. The propagation model includes gaseous absorption as an attenuation term and omits molecular re-radiation. The carrier frequency is a research assumption.

\subsubsection{Element-level spherical wavefronts with tile-wise planar approximation}
Each propagation path between the surface and a point $\bv{o}$ (the BS array centre, a device, an image source, or a scattering point) is a spherical wave. Let $D_q(\bv{o})=\|\bv{p}_q-\bv{o}\|$ and let $\bv{u}_q(\bv{o})=(\bv{o}-\bv{p}_q)/D_q(\bv{o})$ be the unit vector from tile $q$ towards $\bv{o}$. Within one tile we use
\begin{equation}
\|\bv{e}_{q,n}-\bv{o}\|\approx D_q(\bv{o})-\bv{u}_q(\bv{o})^T\bv{r}_n,\label{eq:tileplanar}
\end{equation}
which preserves the spherical phase variation across tile centers. The numerical study evaluates the nominal transfer gains with exact element distances. The neglected second-order term causes a phase error of at most $\pi W^2/(2\lambda_c D_q)$, which stays below $\pi/8$ for $D_q\ge 4W^2/\lambda_c$, that is about 9~cm for $W=2.8$~mm. With $D_{\rm ap}$ denoting the maximum end-to-end aperture dimension, the nominal Rayleigh distance $2D_{\rm ap}^2/\lambda_c$ is about 1.4~m for compact tiles and about 15~m for a pitch of $4W$. Thus, the 2--8~m device links in the compact layout lie beyond its nominal Rayleigh distance, whereas they can lie within it for the spread layout. The BS-array Rayleigh distance is about 8~cm when its diagonal aperture is used, so meter-scale BS-side links are in its far field.

We define the tile response $\bv{a}_q(k,\bv{u})=\big[e^{jk\bv{u}^T\bv{r}_1},\dots,e^{jk\bv{u}^T\bv{r}_{S^2}}\big]^T$ and the BS response $\bv{b}(k,\bv{v})=\big[e^{jk\bv{v}^T\tilde{\bv{r}}_1},\dots,e^{jk\bv{v}^T\tilde{\bv{r}}_{N_t}}\big]^T$, where $\tilde{\bv{r}}_t$ is the offset of BS antenna $t$ from the array centre. Both depend on $k_m$, which is what produces beam split.

\subsubsection{Free-space gain of a path}
Following the measured RIS path-loss model~\cite{tang2021pathloss}, the amplitude of a path of length $d$ that leaves or reaches an element at angle $\vartheta$ from the surface normal is
\begin{equation}
\alpha(f,d,\vartheta)=\sqrt{G_{\rm e}F(\vartheta)}\,\frac{c}{4\pi f d}\,e^{-\kappa(f)d/2}\,e^{-j2\pi f d/c},\label{eq:alpha}
\end{equation}
where $G_{\rm e}=4\pi d_e^2/\lambda_c^2=\pi$ is the element gain and $F(\vartheta)=\cos^{q_{\rm e}}\vartheta$ the normalized element pattern, defined in both half-spaces for a STAR element. Here $\vartheta$ is measured from the normal of the relevant half-space and $q_{\rm e}$ sets the angular roll-off. For broadside paths of lengths $d_1$ and $d_2$, the free-space amplitude of the two-hop product is $G_{\rm e}\lambda^2/(16\pi^2d_1d_2)$, where $\lambda=c/f$. Its squared magnitude gives the corresponding power gain before absorption and STAR losses. The normalization follows the element-level RIS link-budget form in~\cite{tang2021pathloss}.

\subsubsection{Factory scattering: specular and diffuse}
Factory surfaces (machine housings, metal walls, ceiling, concrete floor) are large compared with $\lambda_c$, so a single-bounce reflection from surface $c$ with unit normal $\bv{n}_c$ is modelled by the image method: the path is a straight line from the image $\bv{o}'_c$ of the source mirrored in the surface, so its unfolded length enters~\eqref{eq:alpha}. Its amplitude is scaled by the Fresnel coefficient $\Gamma_{\rm F}(\vartheta_c;\epsilon_c)$ for material permittivity $\epsilon_c$ and by the Kirchhoff roughness factor~\cite{piesiewicz2007}
\begin{equation}
\rho_c(f)=\exp\!\Big(-\frac{8\pi^2 f^2\sigma_{h,c}^2\cos^2\vartheta_c}{c^2}\Big),\label{eq:rough}
\end{equation}
where $\sigma_{h,c}$ is the surface height standard deviation and $\vartheta_c$ the incidence angle. At 850~GHz a painted metal surface with $\sigma_h=50~\mu$m and $\vartheta_c=30^\circ$ gives $10\log_{10}(\rho_c^2)\approx-10$~dB at $f_c$, so roughness must be modelled. The power removed from the specular direction, $(1-\rho_c^2)|\Gamma_{\rm F}|^2$, is scattered diffusely. We represent it by $R$ rays from points $\bv{o}_{c,i}$ spread around the specular point with an angular spread of a few degrees, each with amplitude $\sqrt{(1-\rho_c^2)/R}\,|\Gamma_{\rm F}|\,\varsigma_{c,i}$ and $\varsigma_{c,i}\sim\mathcal{CN}(0,1)$. Both $\rho_c$ and the propagation phase change across the band, so the multipath is frequency selective. The general model allows the number of scattering surfaces to be drawn from a Poisson distribution. The numerical evaluation uses the fixed set of candidate surfaces in Table~\ref{tab:params}. The cited factory measurements establish the relevance of industrial multipath at lower frequencies~\cite{wang2025factory,shao2026factory,ju2022factory}. The ray counts and roughness parameters at 850~GHz are scenario assumptions.

\subsubsection{Blockage}
The LoS link between the surface and device $k$ exists with the 3GPP InF probability~\cite{3gpp38901}
\begin{equation}
\Pr\{\text{LoS}\}=\exp\!\Big(-\frac{d_{2{\rm D}}}{k_{\rm sub}}\Big),\quad k_{\rm sub}=-\frac{d_{\rm clutter}}{\ln(1-r_{\rm cl})},
\end{equation}
where $d_{2{\rm D}}$ is the horizontal separation, $r_{\rm cl}$ is the clutter density and $d_{\rm clutter}$ the typical clutter size. Both the surface (4~m) and the devices (1.5~m) are below the clutter height of a dense factory (6~m), so we use the dense-clutter, low-height (InF-DL) form; this is a 3GPP-inspired extrapolation to 850~GHz, not a validated THz model. A blocked LoS path is removed in the model to represent strong attenuation through metal machinery. Reflected paths are blocked independently with the same law, applied to each segment. The BS--surface link is a planned infrastructure link and has LoS. The direct BS--device link of $k\in\mathcal{K}_{\rm R}$ uses the same InF-DL law (the BS, at 5~m, is also below the clutter height), and it is absent for $k\in\mathcal{K}_{\rm T}$.

\subsubsection{Cascaded channel}
Let $\mathcal{I}$ index the BS-side paths (LoS from the BS, images of the BS, diffuse points), with source $\bv{o}_i$ and scalar gain $c_i$ (one for LoS, the Fresnel, roughness and $\varsigma$ factors otherwise). The BS--tile channel $\bv{G}_q[m]\in\mathbb{C}^{S^2\times N_t}$ on subcarrier $m$ is
\begin{equation}
\bv{G}_q[m]=\sum_{i\in\mathcal{I}} c_i[m]\,\alpha_{q,i}[m]\,\bv{a}_q\big(k_m,\bv{u}_{q,i}\big)\,\bv{b}\big(k_m,\bv{v}_{q,i}\big)^T,\label{eq:G}
\end{equation}
where $\alpha_{q,i}[m]=\alpha\big(f_m,D_q(\bv{o}_i),\vartheta_{q,i}\big)\sqrt{G_{\rm B}}$ includes the BS antenna gain $G_{\rm B}$, $\bv{u}_{q,i}=\bv{u}_q(\bv{o}_i)$, and $\bv{v}_{q,i}$ is the unit vector from $\bv{o}_i$ to tile $q$ (for images, from the reflection point, in the BS's own frame). Likewise, with device-side paths $j\in\mathcal{J}_k$ (LoS to device $k$ if unblocked, images of the device, diffuse points), the tile--device channel is
\begin{equation}
\bv{h}_{k,q}[m]=\sum_{j\in\mathcal{J}_k} c_{k,j}[m]\,\alpha_{q,kj}[m]\,\bv{a}_q\big(k_m,\bv{w}_{q,kj}\big)\in\mathbb{C}^{S^2},\label{eq:h}
\end{equation}
Here $c_{k,j}[m]$ is the device-side path coefficient, $\bv{w}_{q,kj}$ is the unit vector from tile $q$ toward the device, its image, or a scattering point, and $\alpha_{q,kj}[m]$ is the corresponding propagation amplitude including the factor $\sqrt{G_{\rm U}}$ for device antenna gain $G_{\rm U}$. For a diffuse path, its delay includes the segment from the scattering point to the receiver. The end-to-end channel of device $k$, a row vector applied as $y=\bv{g}_k^T\bv{x}$, is
\begin{equation}
\bv{g}_k^T[m]=\sum_{q=1}^{Q}\bv{h}_{k,q}^T[m]\,\bv{\Theta}_q^{\chi_k}[m]\,\bv{G}_q[m]+\bv{g}_{{\rm d},k}^T[m].\label{eq:casc}
\end{equation}
The vector $\bv{g}_{{\rm d},k}[m]$ represents the direct BS--device path and is zero when this path is absent. The total propagation delay through a tile is the sum of the two hop delays. In addition to free-space spreading, the cascade includes frequency-dependent propagation phases, atmospheric attenuation, roughness, and STAR coefficients.

\subsection{Hybrid RSMA Transmission}
We use one-layer RSMA~\cite{mao2018rsma}. The message of device $k$ is split into a common part and a private part. The common parts of all devices are combined and encoded into one common stream $s_c$, and the private parts into private streams $s_1,\dots,s_K$. The common message is encoded jointly across subcarriers into one codeword, because every device must decode it from the whole band. On subcarrier $m$ the BS transmits
\begin{equation}
\bv{x}[m]=\bv{F}_{\rm RF}\big(\bv{f}_c[m]s_c[m]+\textstyle\sum_{k}\bv{f}_k[m]s_k[m]\big)=\bv{W}[m]\bv{s}[m],
\end{equation}
where $\bv{s}[m]=[s_c[m],s_1[m],\ldots,s_K[m]]^T$ collects independent unit-power symbols, and $\bv{F}_{\rm RF}\in\mathbb{C}^{N_t\times N_{\rm RF}}$ is a frequency-flat fully connected phase-shifter network with $|[\bv{F}_{\rm RF}]_{t,r}|=1/\sqrt{N_t}$, $\bv{F}_{\rm D}[m]=[\bv{f}_c[m],\bv{f}_1[m],\dots,\bv{f}_K[m]]$ is the digital precoder ($N_{\rm RF}\ge K+1$), and $\bv{W}[m]=\bv{F}_{\rm RF}\bv{F}_{\rm D}[m]=[\bv{w}_c[m],\bv{w}_1[m],\dots,\bv{w}_K[m]]$. The conducted BS power across all frequency samples obeys $\sum_m\|\bv{W}[m]\|_F^2\le P_{\max}$, where $P_{\max}$ is the transmit-power budget.

\begin{remark}[Beam split at the BS]
Beam split scales with the aperture in wavelengths times the fractional bandwidth. For the $16\times16$ BS array, $16\times B/f_c=0.56<1$ and the edge-of-band pointing error is a fraction of the beamwidth, so a phase-shifter network suffices at the BS. True-time-delay units~\cite{dai2022dpp} would matter only for much larger BS arrays. Beam split is instead governed by the large and reconfigurable surface aperture, where the tile geometry can counter it.
\end{remark}

Device $k$ receives
\begin{equation}
y_k[m]=\bv{g}_k^T[m]\bv{W}[m]\bv{s}[m]+\nu_k[m]+n_k[m],
\end{equation}
where $n_k[m]\sim\mathcal{CN}(0,\sigma^2)$ with sample noise power $\sigma^2=N_0F_{\rm N}B/M$, noise power spectral density $N_0$, and linear receiver noise factor $F_{\rm N}$, and $\nu_k[m]$ is the aggregate transceiver distortion (power-amplifier non-linearity, oscillator phase noise after common-phase-error correction, IQ imbalance), modelled as independent Gaussian noise with variance $\varkappa^2\sum_{j}|\bv{g}_k^T[m]\bv{w}_j[m]|^2$, where $\varkappa$ is the error-vector magnitude~\cite{bjornson2014hwi}. The error-vector magnitude is an assumed aggregate parameter. The model does not resolve oscillator phase noise across time or tones. Each device first decodes $s_c$, treating all private streams as noise, removes it by successive interference cancellation (SIC), and then decodes its own private stream.

\subsection{What the BS and the Devices Know}\label{sec:csi}
Estimating the cascaded channel of several thousand elements on every subcarrier is impractical within a 70~$\mu$s coherence time. The BS therefore holds parametric CSI, as is common for near-field THz systems~\cite{cui2022nf}: the BS--surface geometry is calibrated, and device-side paths are described by estimated locations and complex gains. We call this information the estimated state $\hat{\bv{\Xi}}$ (nominal tile positions $\bv{P}$, estimated device locations $\hat{\bv{t}}_k$, estimated path gains, the blockage pattern), and the BS reconstructs~\eqref{eq:casc} from it. The true channel is random given $\hat{\bv{\Xi}}$, for four reasons:
\begin{enumerate}
\item Actuator error: tile $q$ settles at $\bv{p}_q+\bv{\delta}_q^{\rm a}$, $\bv{\delta}_q^{\rm a}\sim\mathcal{N}(\bv{0},\bv{\Sigma}_{\rm a})$, independent across tiles.
\item Vibration: machines excite a zero-mean stationary displacement $\bv{\delta}^{\rm v}_q(t)$ at time $t$, mostly along the surface normal. It splits into a part common to all tiles (the frame moves rigidly, covariance $\bv{\Sigma}_{\rm v,c}$) and an independent part per tile ($\bv{\Sigma}_{\rm v,i}$). Without position sensing, the whole displacement is unknown to the BS. If each tile has an accelerometer or encoder that is read when the CSI is formed, the residual error after a latency $\Delta t$ has covariance $2\big(1-r_{\rm v}(\Delta t)\big)\bv{\Sigma}_{\rm v}$, where $\bv{\Sigma}_{\rm v}$ denotes the relevant vibration covariance and $r_{\rm v}$ is the normalized autocorrelation of the vibration (for a narrowband vibration at $f_{\rm v}$, $r_{\rm v}(\Delta t)=\cos 2\pi f_{\rm v}\Delta t$).
\item Device-location error: the true location is $\bv{t}_k=\hat{\bv{t}}_k+\bv{\varepsilon}_k$, $\bv{\varepsilon}_k\sim\mathcal{N}(\bv{0},\sigma_{\rm u}^2\bv{I}_3)$, with an assumed standard deviation $\sigma_{\rm u}$ on the order of millimeters.
\item Residual gain error: each device-side path gain is $c_{k,j}(1+\epsilon_{k,j})$, $\epsilon_{k,j}\sim\mathcal{CN}(0,\sigma_{\rm g}^2)$, independent across paths.
\end{enumerate}

For receiver-side channel tracking, the BS sends $n_{\rm p}$ precoded pilot symbols per stream on each frequency sample, at the stream's own power, and device $k$ observes $y^{\rm p}_{kj}[m]=a_{kj}[m]+e_{kj}[m]$ for each of its effective gains $a_{kj}[m]=\bv{g}_k^T[m]\bv{w}_j[m]$, $j\in\{c\}\cup\mathcal{K}$, with $e_{kj}\sim\mathcal{CN}(0,\sigma_{\rm p}^2)$ and $\sigma_{\rm p}^2=\sigma^2/n_{\rm p}$. With the prior power $\pi_{kj}[m]=\E|a_{kj}[m]|^2$, the device forms a zero-intercept linear minimum mean-square-error (LMMSE) estimate $\hat{a}_{kj}=\frac{\pi_{kj}}{\pi_{kj}+\sigma_{\rm p}^2}y^{\rm p}_{kj}$, whose residual $\tilde{a}_{kj}=a_{kj}-\hat{a}_{kj}$ is orthogonal to $\hat{a}_{kj}$ in the uncentered second-moment sense under the assumed pilot model and has mean-square error $\varepsilon_{kj}=\pi_{kj}\sigma_{\rm p}^2/(\pi_{kj}+\sigma_{\rm p}^2)$. For a known nonzero prior mean, an affine LMMSE estimator could also exploit that mean. The simulations compute the prior power from the design channel moments and use independent channel draws for evaluation. A phase rotation common to all paths leaves channel power unchanged and can be tracked when the pilots are informative and the phase is stable over the slot. Differential phase errors change how the paths combine.

\subsection{Geometric Uncertainty: Closed-Form Moments}
Index the propagation paths of device $k$ by $\ell$: a cascaded path $\ell=(q,i,j)$ through tile $q$, BS-side path $i$ and device-side path $j$, or the direct path $\ell={\rm d}$. Displacements of tens of micrometres and location errors of a few millimetres are small compared with link distances of metres, so to first order the phase change of path $\ell$ on frequency $f_m$ is
\begin{equation}
\varphi_{\ell}[m]=k_m\big(\bv{\omega}_{\ell}^T\bv{\delta}_{q_\ell}-\bv{\zeta}_{\ell}^T\bv{\varepsilon}_k\big),\qquad\bv{\delta}_q=\bv{\delta}_q^{\rm a}+\bv{\delta}_q^{\rm v},\label{eq:phase}
\end{equation}
where the tile sensitivity $\bv{\omega}_\ell$ and the location sensitivity $\bv{\zeta}_\ell$ depend on the path type:
\begin{itemize}
\item LoS to the device: $\bv{\omega}_\ell=\bv{u}_{q,i}+\bv{w}_{q,k}$ and $\bv{\zeta}_\ell=\bv{w}_{q,k}$, where $\bv{u}_{q,i}$ and $\bv{w}_{q,k}$ are the unit vectors from tile $q$ towards the BS-side source (BS, its image, or a scattering point) and towards the device.
\item Specular reflection on surface $c$: the device image is $\bv{t}'=\bv{M}_c\bv{t}+2(\bv{n}_c^T\bv{o}_c)\bv{n}_c$ with the mirror matrix $\bv{M}_c=\bv{I}-2\bv{n}_c\bv{n}_c^T$, so a location error $\bv{\varepsilon}_k$ moves the image by $\bv{M}_c\bv{\varepsilon}_k$. With $\bv{w}'_{q}$ the unit vector from tile $q$ to the image, $\bv{\omega}_\ell=\bv{u}_{q,i}+\bv{w}'_{q}$ and $\bv{\zeta}_\ell=\bv{M}_c\bv{w}'_{q}$.
\item Diffuse ray via the fixed point $\bv{o}_{c,i'}$: the path has two segments, tile--point and point--device. With $\bv{w}_{q}^{\rm s}$ the unit vector from tile $q$ to the point and $\bv{\nu}$ the unit vector from the point to the device, $\bv{\omega}_\ell=\bv{u}_{q,i}+\bv{w}_{q}^{\rm s}$ and $\bv{\zeta}_\ell=\bv{\nu}$, which is the same for all tiles.
\item Direct path ($k\in\mathcal{K}_{\rm R}$): $\bv{\omega}_{\rm d}=\bv{0}$ and $\bv{\zeta}_{\rm d}$ is the unit vector from the BS to the device.
\end{itemize}
Amplitude changes are of relative order $10^{-3}$ or less and are neglected. Write $\bv{g}_k[m]=\sum_\ell\bar{\bv{g}}_{k,\ell}[m]\,e^{j\varphi_\ell[m]}(1+\epsilon_\ell)$, where $\bar{\bv{g}}_{k,\ell}$ is the nominal contribution reconstructed from $\hat{\bv{\Xi}}$ and $\epsilon_\ell=\epsilon_{k,j_\ell}$ is shared across all cascaded paths with the same device-side path index $j_\ell$. A direct path, when modeled with gain uncertainty, is assigned its own independent index.

\begin{proposition}\label{prop:moments}
Under the linearized phase model~\eqref{eq:phase} with Gaussian errors, the conditional mean and correlation matrix of the channel of device $k$ given $\hat{\bv{\Xi}}$ are
\begin{align}
\bar{\bv{g}}_k[m]&\triangleq\E[\bv{g}_k[m]\,|\,\hat{\bv{\Xi}}]=\sum_{\ell}\bar{\bv{g}}_{k,\ell}[m]\,e^{-\frac{1}{2}V_{\ell}[m]},\label{eq:mean}\\
\bv{R}_k[m]&\triangleq\E[\bv{g}_k^*[m]\bv{g}_k^T[m]\,|\,\hat{\bv{\Xi}}]\nonumber\\&=\sum_{\ell,\ell'}\bar{\bv{g}}_{k,\ell}^*\bar{\bv{g}}_{k,\ell'}^T\,e^{-\frac{1}{2}V_{\ell\ell'}[m]}\,\big(1+\sigma_{\rm g}^2\mathbb{1}_{j_\ell=j_{\ell'}}\big),\label{eq:corr}
\end{align}
where the sums run over all paths, including the direct path, $V_\ell=\mathrm{Var}(\varphi_\ell)$ and $V_{\ell\ell'}=\mathrm{Var}(\varphi_\ell-\varphi_{\ell'})$:
\begin{align}
V_{\ell\ell'}=k_m^2\Big[&\bv{\omega}_{\ell}^T\bv{\Sigma}_{q_\ell}\bv{\omega}_{\ell}+\bv{\omega}_{\ell'}^T\bv{\Sigma}_{q_{\ell'}}\bv{\omega}_{\ell'}-2\,\bv{\omega}_{\ell}^T\bv{C}_{q_\ell q_{\ell'}}\bv{\omega}_{\ell'}\nonumber\\
&+\sigma_{\rm u}^2\|\bv{\zeta}_{\ell}-\bv{\zeta}_{\ell'}\|^2\Big],\label{eq:V}
\end{align}
$V_\ell=k_m^2\big[\bv{\omega}_\ell^T\bv{\Sigma}_{q_\ell}\bv{\omega}_\ell+\sigma_{\rm u}^2\|\bv{\zeta}_\ell\|^2\big]$, $\bv{\Sigma}_q=\bv{\Sigma}_{\rm a}+\bv{\Sigma}_{\rm v,i}+\bv{\Sigma}_{\rm v,c}$ is the displacement covariance of a tile, and $\bv{C}_{qq'}=\bv{\Sigma}_q$ if $q=q'$ and $\bv{\Sigma}_{\rm v,c}$ otherwise is the cross-covariance of two tiles. Terms with $\bv{\omega}_\ell=\bv{0}$ (the direct path) omit the tile terms, since there is no tile index for that path.
\end{proposition}
\begin{proof}
Suppress the frequency index and condition on $\hat{\bv{\Xi}}$. The random phase in~\eqref{eq:phase} is a linear combination of zero-mean Gaussian displacements and position errors. Independence of the device-location and tile errors gives $V_\ell$. Applying the covariance identity to the phase difference gives~\eqref{eq:V}, including the covariance between the two tiles due to their common vibration. The Gaussian characteristic function gives $\E[e^{j\varphi_\ell}]=e^{-V_\ell/2}$ and $\E[e^{j(\varphi_{\ell'}-\varphi_\ell)}]=e^{-V_{\ell\ell'}/2}$. The independent gain errors have $\E[1+\epsilon_\ell]=1$ and $\E[(1+\epsilon_\ell)^*(1+\epsilon_{\ell'})]=1+\sigma_{\rm g}^2\mathbb{1}\{j_\ell=j_{\ell'}\}$. Substituting these factors into the path sum and its outer product gives~\eqref{eq:mean} and~\eqref{eq:corr}.
\end{proof}

A phase rotation common to all paths affects the conditional mean through $V_\ell$ and cancels from the second moment through $V_{\ell\ell'}$. Pilot tracking can recover this rotation when the phase is stable within a block. Differential phase errors change how the paths combine, which motivates using $\bv R_k[m]$ in the precoder design.

For paths through distinct tiles with independent normal vibration of variance $\sigma_z^2$ and approximately equal normal sensitivities $\omega_z$, the cross term in the channel second moment is attenuated by $\exp(-k_m^2\sigma_z^2\omega_z^2)$. The normal direction cosines add on the reflection side and subtract on the transmission side, so the sensitivity depends on the incidence and departure geometry. Common frame motion cancels from the differential phase between paths with identical tile-displacement sensitivities. It changes the relative phase between a direct path and a surface path with nonzero sensitivity. The location-error term in~\eqref{eq:V} also depends on the difference between device-side direction vectors, which can increase as the tile spacing grows. The full-system evaluation uses exact element distances for the nominal paths and the tile-level linearization in~\eqref{eq:phase} for the residual perturbations.

\subsection{Link-Rate Model and Scheduled-Rate Outage}\label{sec:rates}
For one channel draw, we approximate the pilot-tracked SINRs of device $k$ at frequency sample $m$ as follows. The denominators use the unconditional pilot error powers and the draw-specific distortion power, so these expressions serve as link-level metrics rather than conditional decoding bounds.
\begin{align}
\gamma_{c,k}[m]&=\frac{|\hat{a}_{kc}|^2}{\sum_{j\in\mathcal{K}}|\hat{a}_{kj}|^2+\varkappa^2\sum_{j}|a_{kj}|^2+\sum_j\varepsilon_{kj}+\sigma^2},\label{eq:sinrc}\\
\gamma_{k}[m]&=\frac{|\hat{a}_{kk}|^2}{\sum_{j\in\mathcal{K}\setminus\{k\}}|\hat{a}_{kj}|^2+\varkappa^2\sum_{j}|a_{kj}|^2+\sum_j\varepsilon_{kj}+\sigma^2},\label{eq:sinrp}
\end{align}
where the sums over $j$ without a set include the common stream, and the term $\varepsilon_{kc}$ in~\eqref{eq:sinrp} is the SIC residual caused by the imperfect estimate $\hat{a}_{kc}$. Eq.~\eqref{eq:sinrp} assumes that the common stream has been decoded and removed; when it has not, the private stream is counted as failed (mandatory SIC, see~\eqref{eq:outage}). The $M$ frequencies are quadrature samples of the band; they are not physical OFDM tones. The band-averaged rates are $\bar{R}_{c,k}=\frac{1}{M}\sum_m\log_2(1+\gamma_{c,k}[m])$ and $\bar{R}_k=\frac{1}{M}\sum_m\log_2(1+\gamma_k[m])$, and these raw link rates are converted to net rates with the effective bandwidth $B_{\rm eff}=B(1-o_{\rm CP})(1-o_{\rm p})$. Here $o_{\rm CP}$ and $o_{\rm p}$ denote the cyclic-prefix and pilot fractions. The assumed subcarrier spacing is 3.84~MHz. The numerology uses a useful symbol duration of approximately 260~ns and a 30~ns cyclic prefix. For a 70~$\mu$s coherence block of 241 symbols, $o_{\rm CP}=10.3\%$ and $o_{\rm p}=n_{\rm str}n_{\rm p}/241$ for a full-block slot with $n_{\rm str}$ active streams. Because the common codeword spans all subcarriers, device $k$ decodes it whenever the scheduled common rate satisfies $R_c\le\bar{R}_{c,k}$. This band-wide condition, with the minimum over users taken after averaging over frequency, is less restrictive than requiring decodability on each subcarrier.

The BS fixes, from $\hat{\bv{\Xi}}$, the common rate $R_c=\sum_kC_k$ with portions $C_k\ge0$ and private rates $\tilde{R}_k$. With mandatory SIC, every device must decode the common stream whenever one is scheduled, even if its own portion $C_k$ is zero, because its private SINR~\eqref{eq:sinrp} presumes the common stream has been removed. Device $k$ is therefore in outage in a channel draw if
\begin{equation}
\mathcal{O}_k=\big\{R_c>0\ \text{and}\ R_c>\bar{R}_{c,k}\big\}\cup\big\{\tilde{R}_k>0\ \text{and}\ \tilde{R}_k>\bar{R}_k\big\}.\label{eq:outage}
\end{equation}
(A receiver that falls back to decoding its private stream with the common stream as interference would fail less often; we keep the stricter definition.) The delivered rate is $(C_k+\tilde{R}_k)\mathbb{1}\{\mathcal{O}_k^{\rm c}\}$.

The BS forms its design surrogate from the channel correlation matrices in Proposition~\ref{prop:moments}. It uses the same LMMSE pilot model as the evaluation. With $\bv{R}_k=\bv{R}_k[m]$, the expected stream powers $\pi_{kj}=\bv{w}_j^H\bv{R}_k\bv{w}_j$, the estimation errors $\varepsilon_{kj}=\pi_{kj}\sigma_{\rm p}^2/(\pi_{kj}+\sigma_{\rm p}^2)$ and $\mathcal{D}_k=\varkappa^2\sum_j\pi_{kj}+\sigma^2$ denoting aggregate distortion and noise power, we use the following surrogate SINRs. The signal term is the expected estimated desired-stream power. Each interference term retains the full interfering stream power, providing an additional margin relative to using $\E[|\hat a_{kj}|^2]=\pi_{kj}-\varepsilon_{kj}$ in its place.
\begin{align}
\tilde{\gamma}_{c,k}[m]&=\frac{\pi_{kc}-\varepsilon_{kc}}{\sum_{j\in\mathcal{K}}\pi_{kj}+\sum_j\varepsilon_{kj}+\mathcal{D}_k},\label{eq:surr}\\
\tilde{\gamma}_{k}[m]&=\frac{\pi_{kk}-\varepsilon_{kk}}{\sum_{j\in\mathcal{K}\setminus\{k\}}\pi_{kj}+\sum_j\varepsilon_{kj}+\mathcal{D}_k},\nonumber
\end{align}
The common and private surrogate spectral efficiencies $\tilde{R}_{c,k}$ and $\tilde{R}_k^{\rm s}$ are obtained by averaging $\log_2(1+\tilde{\gamma})$ over the $M$ samples. This deterministic approximation supplies the candidate-slot rates for design. We evaluate achieved link rates and decoding outage separately on held-out channel draws using~\eqref{eq:sinrc}--\eqref{eq:outage}. Section~\ref{sec:solution} handles the dependence of the pilot-error powers on the precoders.

\begin{remark}[Packet size and blocklength]
For a device with payload $D_k$ bits and deadline $T_k$, a dedicated link with effective bandwidth $B_{\rm eff}$ would require spectral efficiency $D_k/(B_{\rm eff}T_k)$. With time sharing, the service accumulated across the slots available before the deadline must instead cover the payload. The normal approximation~\cite{polyanskiy2010} gives a per-channel-use rate reduction of approximately $\sqrt{V/n_k}\,\mathcal{Q}^{-1}(\epsilon_k)$ for blocklength $n_k$ and dispersion $V$, where $\mathcal{Q}$ is the standard Gaussian tail function. The case study uses a swept minimum spectral efficiency and measures link and service outage under the assumed rate model. It does not evaluate end-to-end packet latency or finite-blocklength error probabilities.
\end{remark}

\section{Problem Formulation}\label{sec:problem}
We seek to maximize the weighted sum of scheduled downlink spectral efficiencies while limiting decoding outage for every scheduled device and providing minimum scheduled spectral efficiencies to the devices with service requirements. The design couples the hybrid BS precoders, the discrete STAR states, and the allocation of common and private rates. It also operates on two timescales. Candidate electronic configurations and the frame schedule are computed for the estimated state, whereas tile positions are chosen over a longer interval.

We first consider a single slot for fixed tile positions $\bv{P}$ and estimated state $\hat{\bv{\Xi}}$. Let $\mu_k\ge0$ denote the priority of device $k$. Let $\mathcal{V}=\{\bv{F}_{\rm RF},\{\bv{F}_{\rm D}[m]\}_{m=1}^{M},\{s_{q,n}\},R_c,\{C_k,\tilde{R}_k\}_{k\in\mathcal{K}}\}$ collect the electronic design and scheduled-rate variables. Let $\epsilon_k^{\rm dec}\in[0,1)$ be the decoding-outage target for each device $k\in\mathcal K$. A minimum-rate device $k\in\mathcal K_{\rm G}$ also has a scheduled spectral-efficiency requirement $R_k^{\min}$ and uses its service-profile outage target $\epsilon_k^{\rm dec}=\epsilon_k$. With the outage event $\mathcal{O}_k$ defined in~\eqref{eq:outage}, the single-slot problem is
\begin{subequations}\label{eq:P1cc}
\begin{align}
\max_{\mathcal{V}}\ \ & \sum_{k\in\mathcal{K}}\mu_k\big(C_k+\tilde{R}_k\big)\\
\text{s.t.}\ \ & \Pr\{\mathcal{O}_k\,|\,\hat{\bv{\Xi}}\}\le\epsilon_k^{\rm dec},\quad k\in\mathcal{K},\label{c:cc}\\
& C_k+\tilde{R}_k\ge R_k^{\min},\quad\forall k\in\mathcal{K}_{\rm G},\\
& R_c=\textstyle\sum_kC_k,\ C_k\ge0,\ \tilde{R}_k\ge0,\quad\forall k,\\
& \textstyle\sum_m\|\bv{F}_{\rm RF}\bv{F}_{\rm D}[m]\|_F^2\le P_{\max},\\
& |[\bv{F}_{\rm RF}]_{t,r}|=1/\sqrt{N_t},\quad\forall t,r,\\
& s_{q,n}\in\mathcal{S},\quad\forall q,n,\label{c:star}
\end{align}
\end{subequations}
The first constraint limits the conditional decoding-outage probability for every device, including best-effort devices. A device with no scheduled bits still decodes the common stream when it belongs to the slot and the common stream is active. The second constraint enforces scheduled service for devices with a minimum-rate target. The remaining constraints specify the rate split and the admissible BS power, analog precoder, and STAR states. The numerical study calibrates rates using empirical mean link outage.

Blocked links can make the minimum-rate constraints infeasible, and evaluating the conditional chance constraints repeatedly is computationally expensive. To obtain a tractable design criterion that penalizes unmet service, we replace the conditional chance constraints~\eqref{c:cc} with rate back-off from the surrogate. First, each scheduled rate is the surrogate rate minus a back-off, and a stream whose back-off uses up its surrogate rate is switched off: with stream-activity indicators $\bv{o}=(o_c,(o_k)_{k\in\mathcal K})$ taking values in $\{0,1\}$, $R_c\le o_c\,[\min_k\tilde{R}_{c,k}-\Delta_c]^+$ and $0\le\tilde{R}_k\le o_k\,[\tilde{R}^{\rm s}_k-\Delta_k]^+$, where $[x]^+=\max(x,0)$ and a stream that is off carries no power. Second, the minimum-rate requirement is softened with shortfall variables $\xi_k\ge0$:
\begin{subequations}\label{eq:P1}
\begin{align}
U(\bv{P};\hat{\bv{\Xi}})&=\max_{\mathcal{V},\,\bv{o},\,\bv{\xi}\ge\bv{0}}
\Big[\sum_{k\in\mathcal K}\mu_k(C_k+\tilde{R}_k)\nonumber\\
&\qquad\qquad-\varpi\sum_{k\in\mathcal{K}_{\rm G}}\xi_k\Big]\\
\text{s.t.}\ & \textstyle\sum_{k'}C_{k'}=R_c\le o_c[\tilde{R}_{c,k}-\Delta_c]^+,\,\forall k,\label{c:common}\\
& 0\le\tilde{R}_k\le o_k[\tilde{R}^{\rm s}_k-\Delta_k]^+,\ C_k\ge0,\ \forall k,\label{c:priv}\\
& C_k+\tilde{R}_k\ge R_k^{\min}-\xi_k,\quad\forall k\in\mathcal{K}_{\rm G},\label{c:qos}\\
& \text{power, unit-modulus, codebook constraints},
\end{align}
\end{subequations}
The penalty $\varpi>0$ sets the weight of rate shortfall relative to the scheduled sum rate. The optimizer can accept a shortfall if satisfying a target requires a larger loss in weighted throughput. The all-off schedule, with $\xi_k=R_k^{\min}$ for $k\in\mathcal{K}_{\rm G}$, establishes feasibility even when a critical device is blocked. The simulations use a proportional back-off $b\in[0,1]$, equivalent to setting each gap to $(1-b)$ times its corresponding surrogate rate. The numerical calibration controls an empirical average over evaluated device-slot pairs and layouts. The per-device conditional targets in~\eqref{c:cc} define the benchmark problem and can be violated by the calibrated schedule.

To reduce competition among devices with nearly collinear channels, we allocate transmission time across groups of devices. Slot $s$ serves $\mathcal{U}_s\subseteq\mathcal{K}$ with its own precoders and rates $(C_{s,k},\tilde{R}_{s,k})$, which are zero for $k\notin\mathcal{U}_s$. The rates of the devices in $\mathcal{U}_s$ satisfy~\eqref{c:common}--\eqref{c:priv}. Let $\mathcal{L}$ index the candidate configurations. A frame consists of whole coherence blocks, and $\tau_s\ge0$ is the fraction of blocks assigned to configuration $s$, with $\sum_{s\in\mathcal{L}}\tau_s=1$. We optimize continuous fractions, corresponding to a sufficiently long frame. Per-slot STAR operation assigns separate states $\{s_{q,n}^{(s)}\}$ to each slot. Common-state operation uses a single STAR configuration. The continuous time-allocation model uses the utility
\begin{equation}\label{eq:P1ts}
\begin{aligned}
U_{\rm TS}=\max_{\bv{\tau},\,\bv{\xi}\ge\bv{0}}\ &\textstyle\sum_{k}\mu_k\sum_s\tau_s\big(C_{s,k}+\tilde{R}_{s,k}\big)-\varpi\sum_{k\in\mathcal{K}_{\rm G}}\xi_k\\
\text{s.t.}\ &\textstyle\sum_s\tau_s\big(C_{s,k}+\tilde{R}_{s,k}\big)\ge R_k^{\min}-\xi_k,\ \forall k\in\mathcal{K}_{\rm G},\\
&\sum_{s\in\mathcal{L}}\tau_s=1,\quad \tau_s\ge0,\quad s\in\mathcal{L}.
\end{aligned}
\end{equation}
where $\bv{\tau}=(\tau_s)_{s\in\mathcal{L}}$ and $\bv{\xi}=(\xi_k)_{k\in\mathcal{K}_{\rm G}}$. For fixed candidate rates this program is linear. Jointly optimizing all slot rates and states restores the nonconvex design. Singleton slots yield time-division orthogonal multiple access (OMA). A full-set slot yields one-slot RSMA or, when the common stream is absent, space-division multiple access (SDMA).

The numerical evaluation uses one configuration per coherence block and neglects STAR switching time at block boundaries. If the implementation instead packs several slots into a single coherence block of $L$ OFDM symbols, each slot uses the following symbol budget:
\begin{equation}
\begin{aligned}
L_s^{\rm data} &= L_s-n_{\rm p}J_s-G_s\ge0,\quad s\in\mathcal{A},\\
&\sum_{s\in\mathcal{A}}L_s\le L ,
\end{aligned}
\label{eq:timing}
\end{equation}
where $\mathcal{A}$ contains the active slots, $L_s$ is the integer slot duration in symbols, $J_s$ is the number of active streams, and $G_s$ is the guard in symbols for switching into that slot. Each active slot pays for its pilots and guard once. For example, two one-stream slots in a 241-symbol block require 32 pilot symbols (13.3\%) before switching guards. The payload weight is $L_s^{\rm data}/L$, followed by the cyclic-prefix factor. This timing model defines a separate implementation check for intra-block switching.

On the mechanical timescale, we seek a feasible tile placement that maximizes the expected short-term utility, including the shortfall penalty. With $\mathcal{P}$ defined in~\eqref{eq:Pset}, we formulate the placement problem as
\begin{equation}\label{eq:P0}
\max_{\bv{P}\in\mathcal{P}}\ \E_{\hat{\bv{\Xi}}}\big[U_{\rm TS}(\bv{P};\hat{\bv{\Xi}})\big]\ \approx\ \max_{\bv{P}\in\mathcal{P}}\ \frac{1}{T_{\rm s}}\sum_{u=1}^{T_{\rm s}}U_{\rm TS}(\bv{P};\hat{\bv{\Xi}}_u),
\end{equation}
using a sample-average approximation over $T_{\rm s}$ estimated states drawn from the long-term statistics (large-scale geometry, clutter, blockage). Because the shortfall enters $U_{\rm TS}$, a placement that often fails a critical device scores poorly.

We measure outage on held-out layouts and independent conditional error draws. The number of independent layouts limits the precision of cross-layout comparisons, especially for rare outages. The evaluation reports paired intervals across layouts and treats the empirical outage estimates as link-level indicators.

Problem~\eqref{eq:P1}--\eqref{eq:P0} is difficult for four reasons: (i) the rates are non-convex in the precoders and couple all variables; (ii) the analog precoder has unit-modulus constraints; (iii) the STAR states are discrete, and each state affects all subcarriers differently through~\eqref{eq:lorentz}; and (iv) the tile positions enter the moments through the exponentials of Proposition~\ref{prop:moments} and the non-overlap constraints in~\eqref{eq:Pset} are non-convex.

\section{Proposed Solution}\label{sec:solution}
The design approximates~\eqref{eq:P1ts}--\eqref{eq:P0} by constructing candidate slot rates and then optimizing their time fractions. The approach uses separate objectives for the STAR-state search, digital precoding and frame allocation. We first characterize the BS-side subspace to explain the spatial limitation and reduce the dimension of the precoding problem.

\subsection{The BS-Side Subspace}
\begin{proposition}[BS-side spatial subspace]\label{prop:keyhole}
At frequency $f_m$, let $\mathcal B_m$ be the span of the center-directed steering vectors $\{\bv b(k_m,\bv v_i):i\in\mathcal I\}$, with $\bv v_i$ the BS-side direction of path $i$ toward the aperture center. In the BS-side far field, let every tile be within distance $a$ of the aperture center, let the source--tile distances be at least $D$, and let $R_{\rm BS}=\max_t\|\tilde{\bv r}_t\|$. For the direct and image paths, replacing $\bv v_{q,i}$ by $\bv v_i$ incurs an absolute steering-vector error bounded by $2\sqrt{N_t}k_m R_{\rm BS}a/D$ in Euclidean norm. For fixed first scatterers at the BS end, the associated BS-side direction is fixed. If $\bv v_{q,i}=\bv v_i$ for all tiles and paths, every transposed row of $\bv G_q[m]$ and every cascaded channel $\bv g_k[m]-\bv g_{{\rm d},k}[m]$ belongs to $\mathcal B_m$. Otherwise, its projection residual is bounded by the steering-vector error weighted by the magnitudes of its path coefficients.
\end{proposition}
\begin{proof}
For the direct and specular image paths, the direction at the BS follows from the source--tile displacement, up to an orthogonal reflection of the direction for an image path. The difference between two normalized displacements with minimum norm $D$ is at most twice their displacement difference divided by $D$. Thus $\|\bv v_{q,i}-\bv v_i\|\le2a/D$. The inequality $|e^{jx}-e^{jy}|\le|x-y|$, applied elementwise, bounds the steering difference by $\sqrt{N_t}k_mR_{\rm BS}\|\bv v_{q,i}-\bv v_i\|$. Each row in~\eqref{eq:G} is a linear combination of these steering vectors after transposition. Equation~\eqref{eq:casc} forms another linear combination, so the triangle inequality yields the weighted residual bound. With a single dominant BS-side path and negligible direct links, most channel energy can therefore concentrate along one BS direction.
\end{proof}
For the $16\times16$ BS and the $45$-mm movement region about 5~m away, the transverse aperture-to-distance ratio is about 9~mrad. A numerical spanning set built from the BS-side path responses and surviving direct links over a grid in the movement region and a subset of the frequency samples has dimension $r=14$--17 at a relative residual-energy threshold of $10^{-6}$. When a single infrastructure path dominates the BS-side energy, the resulting concentration limits multiplexing through weaker directions and motivates time sharing and per-slot focusing. The compressed channel is $\bv g_k^{\rm B}[m]=\bv P_{\rm B}^T\bv g_k[m]$, where $\bv P_{\rm B}\in\mathbb C^{N_t\times r}$ is an orthonormal basis. Its second moment is $\bv R_k^{\rm B}[m]=\bv P_{\rm B}^H\bv R_k[m]\bv P_{\rm B}$. The fully digital comparison uses precoders of the form $\bv w=\bv P_{\rm B}^*\bv x$. This reduces the dimension from $N_t=256$ to $r$.

\subsection{STAR States}
For a slot with device set $\mathcal{U}$, the states are chosen element by element, in random order and for two sweeps, to maximize $\sum_{k\in\mathcal{U}}\sum_m\log\|\bv{g}_k^{\rm nom}[m]\|^2$ over the codebook $\mathcal{S}$, where $\bv{g}_k^{\rm nom}$ is the channel in~\eqref{eq:casc} evaluated at the nominal state. Changing the state of one element updates each $\bv{g}_k^{\rm nom}[m]$ through one element contribution. Testing all states of an element costs $O(|\mathcal{S}||\mathcal{U}|Mr)$ in the compressed space. The logarithm balances the channel gains of the devices in the slot (proportional fairness of the gains) and the frequency samples, and the search uses the frequency-dependent codebook~\eqref{eq:star} exactly. This metric is a proxy: it does not see the precoders, which are optimized next. With a common STAR state, $\mathcal{U}$ contains all served devices.

\subsection{Analog Precoder}
For the current slot, we compute the $N_{\rm RF}$ dominant eigenvectors of $\sum_{k\in\mathcal{U}}\bv{R}_k^{\rm B}[m_c]/\tr(\bv{R}_k^{\rm B}[m_c])$, with $m_c$ denoting a reference sample nearest $f_c$. We lift these eigenvectors to the precoder domain via $\bv P_{\rm B}^*$ and retain their phases with modulus $1/\sqrt{N_t}$ to form $\bv F_{\rm RF}$. For a full-column-rank $\bv F_{\rm RF}$, let $\bv H_{\rm RF}=(\bv F_{\rm RF}^H\bv F_{\rm RF})^{-1/2}$. The digital matrix is $\bv F_{\rm D}[m]=\bv H_{\rm RF}\bv X[m]$, where the columns of $\bv X[m]$ are the digital stream precoders in the normalized RF coordinates. This normalization gives total power $\sum_m\|\bv X[m]\|_F^2$. In the compressed model $\tilde{\bv R}_k[m]=\bv T^H\bv R_k^{\rm B}[m]\bv T$ with $\bv T=\bv P_{\rm B}^T\bv F_{\rm RF}\bv H_{\rm RF}$. The fully digital benchmark in the same subspace sets $\bv T=\bv I_r$.

\subsection{Digital Precoders and Rates of One Slot}
Write $\pi_{kj}[m]=\bv{x}_j^H[m]\tilde{\bv{R}}_k[m]\bv{x}_j[m]=\|\bv{L}_{km}\bv{x}_j[m]\|^2$ with $\bv{L}_{km}^H\bv{L}_{km}=\tilde{\bv{R}}_k[m]$, and $\varrho_{kj}=\pi_{kj}/(\pi_{kj}+\sigma_{\rm p}^2)$, so that the useful power in~\eqref{eq:surr} is $\varrho_{kj}\pi_{kj}$ and the error $\varepsilon_{kj}=(1-\varrho_{kj})\pi_{kj}$. For fixed $\varrho$, each surrogate SINR is a ratio of a squared norm and a convex quadratic, $\tilde\gamma=\|\bv{b}\|^2/I$, where $\bv{b}=\sqrt{\varrho}\bv{L}\bv{x}$ is the desired-signal vector for the current decoding step and $I>0$ is its interference, distortion and noise denominator in~\eqref{eq:surr}. The Lagrangian dual transform and the quadratic transform~\cite{shen2018fp} give
\begin{equation}
\begin{aligned}
\ln(1+\tilde\gamma)=\max_{\gamma\ge0,\bv{y}}\ \big[&\ln(1+\gamma)-\gamma\\
&+2\sqrt{1+\gamma}\,\Re\{\bv{y}^H\bv{b}\}\\
&-\|\bv{y}\|^2\big(\|\bv{b}\|^2+I\big)\big],
\end{aligned}\label{eq:qt}
\end{equation}
with optimal $\gamma=\|\bv{b}\|^2/I$ and $\bv{y}=\sqrt{1+\gamma}\,\bv{b}/(\|\bv{b}\|^2+I)$. For fixed $(\gamma,\bv{y})$ the right side is concave in the precoders, because $\bv{b}$ is linear and $I$ is a convex quadratic (the cone representation uses auxiliaries $t_{kj}[m]\ge\|\bv{L}_{km}\bv{x}_j[m]\|$ for the quadratic denominator terms). The common and private rate lower bounds $f_{c,k}$ and $f_k$ follow by averaging the transformed expression over frequency and dividing by $\ln 2$. They enter the convex program
\begin{align}
\max_{\{\bv{X}[m]\},\bv{C},\tilde{\bv{R}}}\ &\textstyle\sum_{k\in\mathcal{U}}\log\big(\delta+C_k+\tilde{R}_k\big)\label{eq:slotprog}\\
\text{s.t.}\ &\textstyle\sum_{k}C_k\le f_{c,k},\ \tilde{R}_k\le f_k,\ C_k,\tilde{R}_k\ge0,\,\forall k\in\mathcal{U},\nonumber\\
&\textstyle\sum_m\|\bv{X}[m]\|_F^2\le P_{\max},\nonumber
\end{align}
a convex conic program with second-order and exponential cones, solved using Clarabel~\cite{goulart2026clarabel}. After each solve, $(\gamma,\bv{y})$ and $\varrho$ are updated from the new precoders. Because $\varrho$ is held fixed inside an iteration, the true surrogate is not guaranteed to increase, so we evaluate it after every iteration, keep the best iterate, and stop after four consecutive iterations without a relative gain of $10^{-3}$ (or after 40 iterations). Streams whose power goes to zero are switched off automatically, which realizes the on/off indicators of~\eqref{eq:P1}. SDMA uses the program with the common stream disabled. RSMA is initialized from SDMA with 10\% of the power assigned to a common beam and uses the same stopping rule. The implementation retains the better of this RSMA candidate and the SDMA candidate under the slot surrogate, and allows a common-free solution from the RSMA run to improve the SDMA candidate.

The slot objective is proportional fair with rate offset $\delta=10^{-3}$: an equal-weight sum rate within a slot tends to give all resources to the strongest device, which produces slots that cannot help the minimum-rate devices. The fair slot designs act as candidate rate vectors, and the time-allocation step applies the weighted shortfall-penalized objective of~\eqref{eq:P1ts} to those vectors.

\subsection{Time Fractions}
For a set of candidate slots with surrogate rate vectors $\bv{r}_s=(C_{s,k}+\tilde{R}_{s,k})_k$, problem~\eqref{eq:P1ts} in $(\bv{\tau},\bv{\xi})$ is a linear program. The candidates are every served device alone (single-user focusing with the dominant eigenvector), every pair of one reflection-side and one transmission-side device, and all served devices together, each with RSMA or SDMA. We compare five schemes: OMA (singletons only), SDMA-TS and RSMA-TS (singletons, pairs and the full set, without or with a common stream), and SDMA and RSMA with all devices in one slot and no time sharing. Each is run with per-slot and with common STAR states. The design scales minimum-rate targets by $1/b_{\rm d}$ with $b_{\rm d}=0.7$ to provide nominal headroom. The applied back-off $b$ is calibrated separately as described in Section~\ref{sec:results}. For the max-min comparison, the frame allocation maximizes a scalar $z$ constrained by $z\le\sum_s\tau_s r_{s,k}$ for every reachable device $k$, where $r_{s,k}$ is entry $k$ of the candidate slot rate vector.

\subsection{Tile Positions}
Starting from the compact layout, each tile in random order tries in-plane moves of $\pm W/2$ along $x$ and $y$ and normal moves of $\pm0.5$~mm, subject to~\eqref{eq:Pset}. For each trial the element channels of that tile are recomputed exactly, its STAR states are re-optimized for the devices of every active slot (or for all devices with a common state), and~\eqref{eq:P1ts} is evaluated with the precoders and time fractions held fixed. A move is kept only if the utility increases, so the utility is non-decreasing over the search. After one sweep over the tiles, the STAR states, precoders and time fractions are redesigned at the new positions. Because the moments already average over the small-scale errors and the large-scale geometry is fixed within the long-term interval, we use $T_{\rm s}=1$ in~\eqref{eq:P0}.

\subsection{Complexity and Convergence}
One sweep of the STAR search costs $O(N|\mathcal S||\mathcal U|Mr)$ with elementwise updates in the compressed space. Forming a dense $r\times r$ covariance by explicit path-pair outer products requires $O(M L_k^2 r^2)$ operations per device, where $L_k$ is the total number of cascaded and direct path contributions. The number of stream--device quadratic terms in a slot scales as $O(M|\mathcal U|(|\mathcal U|+1))$. The digital stage records the best iterate under its own surrogate, and the position search accepts moves only when the utility with the fixed slot precoders and time fractions improves. The subsequent redesign can change the frame utility. The procedure provides a feasible candidate design under the relaxed model and carries no stationarity guarantee for the joint problem.

\section{Numerical Results}\label{sec:results}

\subsection{Setup and Metrics}
\begin{table}[!t]
\centering
\caption{Reference simulation parameters and sensitivity settings.}
\label{tab:params}
\footnotesize
\setlength{\tabcolsep}{3pt}
\renewcommand{\arraystretch}{1.08}
\begin{tabular}{@{}p{0.29\columnwidth}p{0.67\columnwidth}@{}}
\toprule
Parameter & Value and sensitivity settings \\
\midrule
Carrier, bandwidth, samples & 850~GHz, 30~GHz (10, 50), $M=16$ \\
Pilots & $n_{\rm p}=16$ per stream and sample \\
Atmosphere & 25~$^\circ$C, 1013~hPa, RH 50\% (70\%) \\
BS & $16\times16$ UPA, $N_{\rm RF}=K+1$, $G_{\rm B}=10$~dBi, 5~m from the surface at $30^\circ$ \\
Devices & $K=6$, 3 per side, 1 minimum-rate device per side, $R^{\min}=0.1$ (0.05--0.4) bit/s/Hz \\
Device antenna & $G_{\rm U}=20$~dBi (aligned with the surface) \\
Surface & $Q=4\times4$ tiles of $16\times16$, $N=4096$, $d_e=\lambda_c/2$ \\
Tile width, region, stroke, gap & $W=2.82$~mm, $A=16W=45$~mm, $Z=1$~mm, $g=0.1$~mm \\
STAR element & passive shared-$\Gamma$, $Q_{\rm e}=10$, $\eta=-1$~dB (assumed), $|\mathcal{S}|=24$ \\
Element pattern & $G_{\rm e}=\pi$, $q_{\rm e}=3$ (assumed) \\
Geometry & surface 4~m, BS 5~m, devices 1.5~m high; devices 2--8~m, $\pm60^\circ$ \\
Clutter & InF-DL $r_{\rm cl}=0.4$, $d_{\rm clutter}=2$~m (DL $r_{\rm cl}=0.6$; SL $r_{\rm cl}=0.2$, 10~m) \\
Scatterers & BS side: floor, metal wall; device side: floor, one machine face, 4 diffuse rays each \\
Roughness & $\sigma_h=100~\mu$m concrete, $30~\mu$m metal \\
Errors & $\sigma_{\rm a}=2~\mu$m; $\sigma_{\rm d}=10$ (40)~$\mu$m; $\sigma_{\rm c}=30$ (100)~$\mu$m; $\sigma_{\rm u}=1$ (5)~mm; $\sigma_{\rm g}^2=0.01$ \\
Transceiver EVM & $\varkappa=-25$~dB (assumed) \\
Noise & $-174$~dBm/Hz, $F_{\rm N}=10$~dB \\
Power & $P_{\max}=30$~dBm (20, 40) \\
Solver & $\varpi=10$, $\delta=10^{-3}$, design factor $b_{\rm d}=0.7$, 40 evaluation draws \\
\bottomrule
\end{tabular}
\end{table}

\begin{table*}[!t]
\centering
\caption{Paired comparisons on held-out layouts. Entries report percentage change relative to the second scheme, mean paired difference with 95\% CI, and the number of layouts favoring the first scheme when available. Per-slot STAR is used unless indicated. The abbreviation n.s. indicates that the CI spans zero.}
\label{tab:pairs}
\footnotesize
\begin{tabular}{lll l}
\toprule
Comparison & Metric & 30 dBm & 40 dBm \\
\midrule
SDMA-TS vs OMA & net sum (Gbit/s) & $+17\%$; $24.7\pm53.1$; 6/10 (n.s.) & $+31\%$; $47.6\pm57.4$; 7/10 (n.s.) \\
SDMA-TS vs OMA & ergodic sum (bit/s/Hz) & $+62\%$; $3.96\pm3.99$; 7/10 (n.s.) & $+79\%$; $5.53\pm4.32$; 8/10 \\
SDMA-TS vs OMA & worst device (bit/s/Hz) & $+22\%$; $0.028\pm0.030$; 10/10 (n.s.) & $+37\%$; $0.17\pm0.20$; 9/10 (n.s.) \\
Time sharing vs one slot (SDMA-TS vs SDMA) & net sum (Gbit/s) & $+36\%$; $44.8\pm19.6$; 10/10 & $+40\%$; $58.0\pm24.7$; 10/10 \\
Time sharing vs one slot (SDMA-TS vs SDMA) & worst device (bit/s/Hz) & $+133\%$; $0.088\pm0.084$; 10/10 & $+88\%$; $0.30\pm0.19$; 10/10 \\
RSMA-TS vs SDMA-TS & ergodic sum (bit/s/Hz) & $-0.1\%$; $-0.008\pm0.015$ & $+0.02\%$; $0.003\pm0.032$ \\
RSMA vs SDMA (one slot) & ergodic sum (bit/s/Hz) & $-0.1\%$; $-0.009\pm0.012$ & $-0.05\%$; $-0.006\pm0.024$ \\
Per-slot vs common STAR (SDMA-TS) & net sum (Gbit/s) & $+5\%$; $7.5\pm12.0$; 9/10 (n.s.) & $+13\%$; $22.9\pm26.5$; 7/10 (n.s.) \\
Per-slot vs common STAR (SDMA-TS) & worst device (bit/s/Hz) & $+31\%$; $0.036\pm0.031$; 10/10 & $+30\%$; $0.15\pm0.09$; 10/10 \\
\bottomrule
\end{tabular}
\end{table*}
\begin{figure*}[!t]
\centering
\subfloat[Sum-rate objective]{\includegraphics[width=0.48\textwidth]{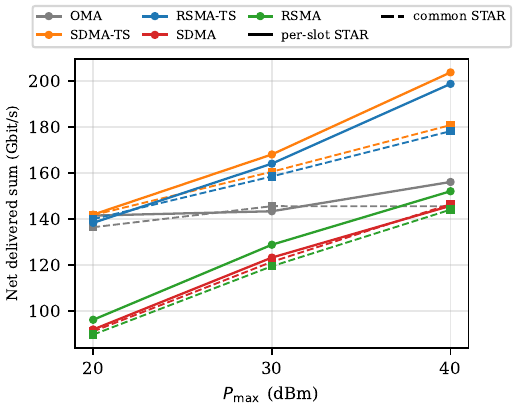}\label{fig:fullpowera}}\hfill
\subfloat[Max-min objective]{\includegraphics[width=0.48\textwidth]{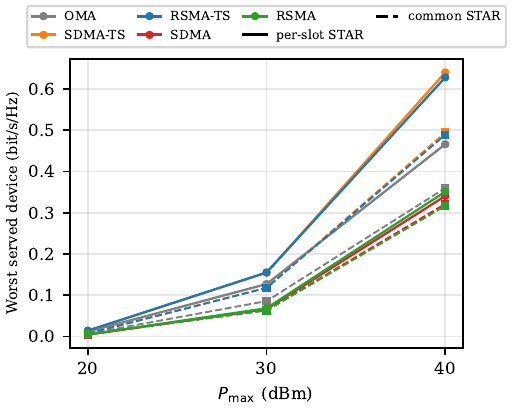}\label{fig:fullpowerb}}
\caption{Full system, held-out realizations, fixed compact tiles. (a) Net delivered sum throughput of the sum-rate schedule; (b) delivered rate of the worst served device under the max-min schedule. Solid: per-slot STAR states; dashed: one common STAR state. The ergodic comparison of RSMA and SDMA is reported in Table~\ref{tab:pairs}.}
\label{fig:fullpower}
\end{figure*}

The design in Section~\ref{sec:solution} was evaluated on 20 factory layouts with the parameters in Table~\ref{tab:params}. Each layout has six devices, three on each side of the partition, including one minimum-rate device on each side. The reference case uses a $16\times16$ BS with seven RF chains, 16 frequency samples, 16 pilots per stream, and InF-DL blockage. A device is served if at least one path to it survives blockage; on average $4.15\pm0.51$ of the six devices are served, and 32.5\% of the minimum-rate devices have no surviving path across the 20 layouts. On average $1.1$ reflection-side devices per realization also have a direct link. For each evaluated candidate slot, we use 40 conditionally independent error draws per layout with pilot-tracked LMMSE receivers and mandatory SIC.

We report four metrics. The ergodic sum rate averages raw spectral efficiencies across channel draws with per-draw rate adaptation. The delivered sum throughput counts a fixed scheduled rate only on successful decoding and deducts the cyclic prefix and a normalized pilot overhead for each slot type. The worst served-device rate takes the minimum, over devices with a surviving path, of the per-device delivered spectral efficiency averaged across draws. The QoS outage counts trials below $R^{\min}$ for minimum-rate devices, including devices with no surviving path.

The reported throughput uses the payload weight $\tau_s(1-J_sn_{\rm p}/L)$ for slot type $s$, with $L=241$ and $n_{\rm p}=16$, followed by $1-o_{\rm CP}$. The pilot count uses one stream for a single-user slot and one private stream per device for a multiplexed slot, with an additional common stream when RSMA schedules a common rate above $10^{-3}$~bit/s/Hz. The worst served-device rate and the minimum-rate outage use delivered spectral efficiency before these overhead deductions.
\begin{figure*}[!t]
\centering
\subfloat[Robust vs nominal design, RSMA, 30~dBm]{\includegraphics[width=0.48\textwidth]{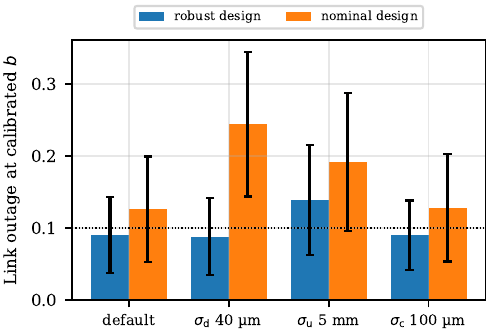}\label{fig:fullrobusta}}\hfill
\subfloat[Bandwidth, humidity and STAR model]{\includegraphics[width=0.48\textwidth]{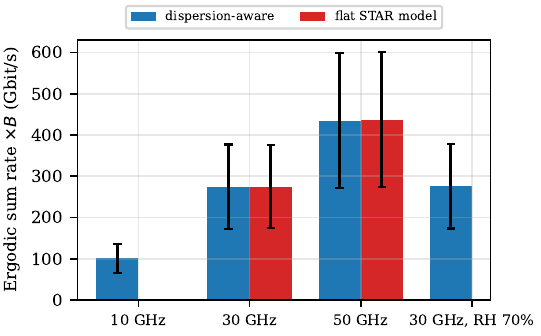}\label{fig:fullrobustb}}
\caption{Common STAR state, all devices in one slot, RSMA, 30~dBm. (a) Mean link outage on the held-out realizations at the calibrated back-off, for designs from the moments of Proposition~\ref{prop:moments} (robust) and from the error-free channel (nominal), at the default errors and with one error increased. (b) Ergodic sum rate times bandwidth over 20 realizations, for designs with the dispersion-aware STAR model and with a flat model (flat model run at 30 and 50~GHz only); 95\% CIs across realizations.}
\label{fig:fullrobust}
\end{figure*}

For each scheme and STAR mode, we choose the largest rate factor $b$ on a grid of step 0.01 with mean link outage at most 10\% at 30~dBm for the sum-rate schedules on layouts 1--10. The grid covers 0.30--1.00, with 0.30 as the fallback if no grid value meets the target. The delivered metrics use layouts 11--20. The calibrated values are 0.87--0.90 for OMA, 0.77--0.79 for SDMA-TS and RSMA-TS, and 0.64--0.68 for one-slot SDMA or RSMA. The comparison reuses these values at the other powers and objectives. Consequently, the results show performance with transferred calibration across conditions. In particular, the measured outage can differ across schemes. A common rate below $10^{-3}$~bit/s/Hz is treated as inactive.

Because layouts differ in the number of reachable devices and the availability of direct links, we report paired differences and 95\% confidence intervals across layouts alongside ratios of the layout means.

\subsection{Multiple Access and Time Sharing}
Figure~\ref{fig:fullpower} and Table~\ref{tab:pairs} summarize the held-out comparisons. With per-slot STAR states, SDMA with time sharing improves delivered sum throughput over one-slot SDMA in all ten held-out layouts at 30 and 40~dBm. The mean throughput improvements are $44.8\pm19.6$~Gbit/s at 30~dBm and $58.0\pm24.7$~Gbit/s at 40~dBm (paired 95\% confidence intervals), corresponding to 36\% and 40\%. Under max-min allocation, the worst served-device advantage is $0.088\pm0.084$ and $0.30\pm0.19$~bit/s/Hz, respectively. These gains are consistent with Proposition~\ref{prop:keyhole}: scheduling avoids requiring several weak and nearly collinear links to share every transmission interval.

At 20~dBm, OMA and both time-sharing schemes deliver about 140~Gbit/s. At 30 and 40~dBm, SDMA-TS delivers about 168 and 204~Gbit/s, versus 143 and 156~Gbit/s for OMA. The paired intervals for these delivered-rate differences include zero. Pilot overhead and the calibrated scheduled-rate factors reduce the gains observed in raw ergodic rates. Each additional active stream adds 16 pilot symbols per full-block slot. Thus the 62--79\% raw ergodic-rate advantage narrows to 17--31\% in delivered sum throughput under the assumed overhead.

The absolute differences in mean ergodic sum rate between RSMA and SDMA are below 0.6\% for the weighted sum-rate and max-min schedules, and below 0.9\% when the proportional-fair schedules are included. The reported rate comparisons characterize the effect of enabling the common stream with the candidate designs and allocations used here. Common-stream use in the final schedule would be measured by the sum of $\tau_s$ over allocated slots with an active common stream.

\subsection{STAR Mode, Hybrid Precoding and Movable Tiles}
Under SDMA-TS with the max-min schedule, per-slot configurations improve the worst served-device rate by $0.036\pm0.031$~bit/s/Hz at 30~dBm and $0.15\pm0.09$~bit/s/Hz at 40~dBm, corresponding to 31\% and 30\% in Table~\ref{tab:pairs}. The delivered sum-throughput gains of 5\% and 13\% have paired confidence intervals spanning zero. Per-slot states allow the surface to focus on a smaller set of intended devices. These gains assume negligible switching latency for the 4096 elements. A positive guard period reduces the payload time as in~\eqref{eq:timing}.

With a common STAR state and one full-set slot, the fully digital benchmark in the retained BS subspace improves the surrogate sum rate by $0.10\pm0.27$~bit/s/Hz (1.0\% on average). This comparison supports the use of seven RF chains for the evaluated geometry and slot design.

The tile search moved an average of 12.0 tiles with per-slot states and 9.8 tiles with a common state, taking about 22~s per layout. The fixed-precoder search utility improved by 0.30\% and 0.61\%, respectively. After redesign, the RSMA-TS surrogate sum rate increased by 0.29\% and 1.1\%, respectively. Across 20--40~dBm, the change in mean delivered sum throughput for RSMA-TS under sum-rate scheduling stays below 1\% in magnitude with per-slot STAR states and ranges from $-1.3\%$ to $+5.5\%$ with a common STAR state. The available stroke adjusts path delays and inter-tile phase, while the electronic states already compensate much of the nominal phase mismatch in these layouts.

\subsection{Robustness, Bandwidth and Humidity}

\begin{figure*}[!t]
\centering
\subfloat[QoS outage versus minimum rate]{\includegraphics[width=0.48\textwidth]{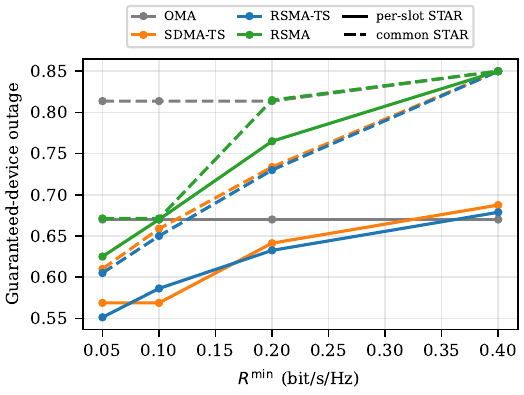}\label{fig:fullqosa}}\hfill
\subfloat[Clutter density, per-slot STAR, 30~dBm]{\includegraphics[width=0.48\textwidth]{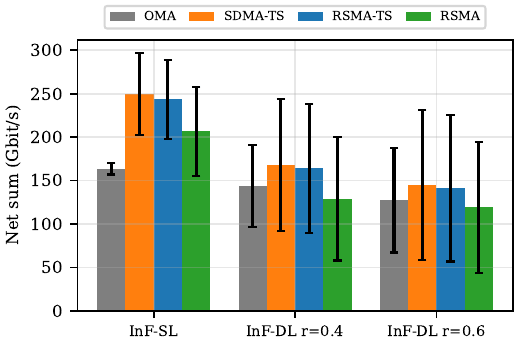}\label{fig:fullqosb}}
\caption{Held-out realizations, 30~dBm, sum-rate schedule. (a) QoS outage of the minimum-rate devices versus $R^{\min}$ (unserved devices count as in outage); solid: per-slot STAR, dashed: common STAR. (b) Net delivered sum throughput under sparse (InF-SL) and dense (InF-DL) clutter, per-slot STAR, with 95\% CIs across realizations.}
\label{fig:fullqos}
\end{figure*}

In Fig.~\ref{fig:fullrobusta}, the moment-based design increases ergodic sum rate by 0.6--1.0\%, with intervals including zero. At the same back-off factor, it yields 9--14\% mean link outage across the tested errors. The nominal design gives 13\% at the default errors, 19\% with 5~mm location uncertainty and 24\% with 40~$\mu$m differential vibration. Each design may schedule a different absolute rate at the same factor, so the comparison measures sensitivity under common calibration. A throughput comparison at matched outage would require a separate calibration for each design.

Figure~\ref{fig:fullrobustb} shows that the ergodic spectral efficiency of RSMA falls from 11.0 to 9.9 and 9.4~bit/s/Hz at $B=10$, 30 and 50~GHz, as the noise power grows with $B$ and beam split widens; the raw throughput rises from 110 to 296 and 471~Gbit/s. With a common STAR state, the net delivered throughput rises from 48 to 146 and 240~Gbit/s for OMA and from 44 to 120 and 191~Gbit/s for RSMA with all devices in one slot under the reported overhead convention.

Increasing relative humidity from 50\% to 70\% changes the rates by less than 1\% in the modeled 850-GHz window. Designing the STAR states with frequency-flat coefficients and evaluating both designs with the dispersive response changes the ergodic result by $-0.06\%$ at 30~GHz and $+0.5\%$ at 50~GHz, with intervals spanning zero. Elements in the same codebook state share a frequency-dependent phase, which a pilot-tracked receiver can follow. Relative phase changes between different states can still affect the array gain.

\subsection{Industrial QoS and Clutter}

In Fig.~\ref{fig:fullqosb}, the mean number of served devices falls from 6.0 of 6 in sparse clutter to 4.15 and 2.45 in the two dense-clutter settings. The SDMA-TS and OMA delivered throughput means are 250 and 164~Gbit/s in sparse clutter, 168 and 143~Gbit/s at $r_{\rm cl}=0.4$, and 145 and 127~Gbit/s at $r_{\rm cl}=0.6$. Fewer surviving links leave fewer useful device pairings, which narrows the observed difference between the two schemes.

As shown in Fig.~\ref{fig:fullqosa}, at 30~dBm and $R^{\min}=0.1$~bit/s/Hz, the QoS outage is 57\% for SDMA-TS with per-slot STAR states and 67--81\% for OMA. The outage curves count unreachable devices as well as failures among reachable devices. The sum-rate objective can leave a weak device with too little scheduled service when the throughput loss exceeds the shortfall penalty. Decoding failure adds a second limitation. At 30~dBm with per-slot STAR states, reusing the sum-rate back-off for the max-min versions of OMA, SDMA-TS, and RSMA-TS yields approximately 11--24\% mean link outage. In the reported comparisons, the QoS effect of per-slot STAR states depends on the rate target and multiple-access scheme. The observed outage combines coverage, time allocation and rate-calibration effects.

\subsection{Convergence and Run Time}
The fractional-programming loop ran $11.4\pm1.8$ iterations per multi-device slot before the four-stall stop, at about 0.8~s per iteration (CVXPY/Clarabel, $M=16$, $r\approx15$). One layout, with all experiments of this section (three powers, two STAR modes, five schemes, three frame objectives, the movable-tile searches and the robustness, bandwidth and QoS variants), took $57\pm16$~min on one CPU core; the 20 realizations ran on four cores in about 5.5~hours.

\section{Conclusion}\label{sec:conclusion}
We studied an 850-GHz IIoT downlink in which a tile-based STAR-RIS serves devices on both sides of a factory partition. The channel analysis identifies a common spatial direction at the BS when the BS--surface hop is dominated by one path. The geometric-error moments distinguish common phase rotation from differential vibration and show how the placement of the BS and the devices affects coherence across tiles. Under the evaluated time-allocation and overhead convention, scheduling device groups and configuring the surface per slot improves delivery to weak served devices. In the full-system comparisons, enabling RSMA has little effect on the ergodic sum rate under the evaluated scheduling policies. For RSMA with time sharing, tile movement changes mean delivered sum throughput by less than 1\% in magnitude with per-slot STAR states and by up to 5.5\% with a common state. The high minimum-rate outage in dense clutter motivates coverage improvements and calibration for each scheduling objective. Measured STAR switching times and outage targets calibrated for each objective would help quantify the gains for specific deployments.

\end{document}